\documentclass[preprint]{aastex}
\usepackage{psfig,natbib}
\received{9 October 2000}  

\newcommand{\citepeg}[1]{\citep[{e.g.,}][]{#1}}
\newcommand{\citepcf}[1]{\citep[{see}\phantom{}][]{#1}}
\newcommand{\rha}[0]{\rightarrow}

\def\lsim{\hbox{ \rlap{\raise 0.425ex\hbox{$<$}}\lower 0.65ex\hbox{$\sim$}}}
\def\gsim{\hbox{ \rlap{\raise 0.425ex\hbox{$>$}}\lower 0.65ex\hbox{$\sim$}}}

\def\arcsec{\hbox{$^{\prime\prime}$}}

\def\ale{\mathrel{\hbox{\rlap{\hbox{\lower4pt\hbox{$\sim$}}}\hbox{$<$}}}}
\def\age{\mathrel{\hbox{\rlap{\hbox{\lower4pt\hbox{$\sim$}}}\hbox{$>$}}}}

\shorttitle{Offset Distribution of Gamma-Ray Bursts}
\shortauthors{Bloom, Kulkarni, \& Djorgovski}
\begin{document}
\title{The Observed Offset Distribution of Gamma-Ray Bursts from Their
       Host Galaxies: A Robust Clue to the Nature of the
       Progenitors\footnotemark\footnotetext{Partially based on
       observations with the NASA/ERA Hubble Space Telescope, obtained
       at the Space Telescope Science Institute, which is operated by
       the Association of Universities for Research in Astronomy,
       Inc.~under NASA contract No.~NAS5-26555.}\phantom{`}${}^{\rm
       ,}$\footnotemark\footnotetext{In addition, some of the data
       presented herein were obtained at the W.~M.~Keck Observatory,
       which is operated as a scientific partnership among the
       California Institute of Technology, the University of
       California and the National Aeronautics and Space
       Administration, and was made possible by the generous financial
       support of the W.~M.~Keck Foundation. 
}}

\author{J.~S.~Bloom, S.~R.~Kulkarni, S.~G.~Djorgovski}
       \affil{Palomar Observatory 105--24, California Institute of
       Technology, Pasadena, CA 91125, USA}

\begin{abstract}

We present a comprehensive study to measure the locations of
$\gamma$-ray bursts (GRBs) relative to their host galaxies.  In total,
we find the offsets of 20 long-duration GRBs from their apparent host
galaxy centers utilizing ground-based images from Palomar and Keck and
space-based images from the Hubble Space Telescope (HST).  We discuss
in detail how a host galaxy is assigned to an individual GRB and the
robustness of the assignment process. The median projected angular
(physical) offset is 0.17 arcsec (1.3 kpc).  The median offset
normalized by the individual host half-light radii is 0.98 suggesting
a strong connection of GRB locations with the UV light of their hosts.
This provides strong observational evidence for the connection of GRBs
to star-formation.

We further compare the observed offset distribution with the predicted
burst locations of leading stellar-mass progenitor models. In
particular, we compare the observed offset distribution with an
exponential disk, a model for the location of collapsars and promptly
bursting binaries (e.g.~helium star--black hole binaries). The
statistical comparison shows good agreement given the simplicity of
the model, with the Kolmogorov-Smirnov probability that the observed
offsets derive from the model distribution of $P_{\rm KS} = 0.45$.  We
also compare the observed GRB offsets with the expected offset
distribution of delayed merging remnant progenitors (black
hole--neutron star and neutron star--neutron star binaries). We find
that delayed merging remnant progenitors, insofar as the predicted
offset distributions from population synthesis studies are
representative, can be ruled out at the $2 \times 10^{-3}$ level. This
is arguably the strongest observational constraint yet against delayed
merging remnants as the progenitors of long-duration GRBs.  In the
course of this study, we have also discovered the putative host
galaxies of GRB 990510 and GRB 990308 in archival HST data.
\end{abstract}

\keywords{astrometry---cosmology: miscellaneous --- cosmology:
          observations --- gamma rays: bursts---methods: statistical}

\section{Introduction}

For some thirty years since the discovery of gamma-ray bursts
\citep[GRBs;][]{kso73}, a basic understanding of the nature of the
brief intense flashes of $\gamma$-rays remained elusive.  Throughout
much of the 1970s and 1980s the prevailing view was that GRBs arise
from the surface of neutron stars in and around our Galaxy
\citep[see][for a review]{lam95}, though, by the mid-1990s, the
isotropic distribution of GRBs on the sky \citepcf{fm95} served as the
cornerstone of mounting evidence suggesting an extra-galactic origin
\citep[see][for a review]{pac95}.  The main impedance to progress was
the difficulty of localizing bursts to an accuracy high enough to
unequivocally associate an individual GRB with some other
astrophysical entity.  In large measure the localization problem was
due to both the transient nature of the phenomena and the fact that
the incident direction of $\gamma$-rays are difficult to pinpoint with
a single detector; for example, the typical 1-$\sigma$ uncertainty in
the location of a GRB using the {\it Burst and Transient Source
Experiment} (BATSE) was $4$--$8$ degree in radius \citep{bpk+99}. The
Interplanetary Network \citep[IPN; see][]{cbb+99} localized GRBs using
burst arrival times at several spacecrafts throughout the Solar System
and provided accurate localizations (3 $\sigma$ localizations of
$\sim$few to hundreds $\times$ arcmin$^2$) to ground-based observers;
however, the localizations were reported with large time delays (days
to months after the GRB).

The crucial breakthrough came in early 1997, shortly following the
launch of the BeppoSAX satellite \citep{bbp+97}. On-board instruments
\citep{fcd+97,jmb+97} were used to rapidly localize the prompt and
long-lived hard X-ray emission of the GRB of 28 February 1997 (GRB
970228) to a 3 $\sigma$ accuracy of 3 arcmin (radius) and relay the
location to ground-based observers in a matter of hours.  Fading X-ray
\citep{cfh+97} and optical \citep{vgg+97} emission, the so-called
``afterglow,'' associated with GRB 970228 were discovered.
Ground-based observers noted \citep{mkd+97,vgg+97} a faint nebulosity
in the vicinity of the optical transient (OT) afterglow. Subsequent
{\it Hubble Space Telescope} (HST) imaging resolved the nebulosity
\citep{slp+97} and showed that the morphology was indicative of a
distant galaxy \citep{slp+97}.  Three years later, we now know the
redshift of this faint, blue galaxy is $z=0.695$ \citep{bdk01}.  

The next prompt localization of a GRB yielded the first confirmed
distance to the GRB through optical absorption spectroscopy: GRB
970508 occurred from a redshift \hbox{$z \ge 0.835$} \citep{mdk+97}.
The first radio afterglow was detected from GRB 970508 which, through
observations of scintillation, led to the robust inference of
superluminal motion of the GRB ejecta \citep{fkn+97}.  These
measurements (along with the dozen other redshifts now associated with
individual GRBs) have effectively ended the distance scale debate and
solidified GRBs as one of the most energetic phenomena known
\citepcf{kbb+00,fks+01}.

The cosmological nature of GRBs now frames our basic understanding of
the physics of GRB phenomena\footnotemark\footnotetext{The emergent
picture described herein is reserved to the so-called long duration
GRBs, those lasting for a duration $\age 2$ sec, since no
short-duration bursts have yet been well-localized on rapid timescales
($\ale $ few days).}.  The general energetics are well-constrained:
given the observed fluences and redshifts, approximately
$10^{51}$--$10^{53}$ erg in $\gamma$-ray radiation is released in a
matter of a few seconds in every GRB. The GRB variability timescale
suggests that this energy is quickly deposited by a ``central
engine'' in a small volume of space (radius of $\sim 30$ km) and is
essentially optically thick to $\gamma$-ray radiation at early times.
This opaque fireball of energy then expands adiabatically and
relativistically until the $\gamma$-ray radiation can escape; the
emitting surface of the GRB is likely to be 10$^{15}$--$10^{17}$ cm
from the explosion site and probably arises from the interaction of
internal shocks initiated by the central engine \citepeg{frrw99}.
Only a small amount of baryonic matter ($\sim 10^{-5} M_\odot$) can be
entrained with the fireball since too much baryonic matter, a
condition referred to as the ``baryonic loading'' problem, would
essentially stall the relativistic expansion of the fireball. The
transient afterglow phenomena is thought to be due to synchrotron
radiation arising from the interaction of the relativistic ejecta and
the ambient medium surrounding the burst site \citep[see][for
reviews]{cpkw00,kbb+00,dfk+01a}. The relativistic nature of the
expanding shock (which also gives rise to the GRB) is required to
avoid the so-called compactness problem \citepcf{pir99} and, as
mentioned above, was observationally confirmed with radio
scintillation measurements of the afterglow of GRB 970508 \citep{fkn+97}.

While the GRB emission and the afterglow phenomenon are now reasonably
well-understood, one large outstanding question remains: what makes a
$\gamma$-ray burst?  Specifically what are the astrophysical objects,
the ``progenitors'', which produce $\gamma$-ray bursts?  To-date
several theoretical considerations appear to implicate the progenitors
as stellar-mass systems involving a compact source, probably a black
hole (BH).  First, the implied (isotropic) energy release in
$\gamma$-rays are typically 10$^{-3}$--$10^{-1}$ times the rest-mass
energy of the Sun.  The estimated efficiency of conversion of the
initial input energy (either Poynting flux or baryonic matter) to
$\gamma$-rays ranges from $\sim$1\% \citepeg{kum99} to as much as
$\sim$60\% \citepeg{ks01}; therefore, the best-guess estimate of the
total energy release is roughly comparable to the rest-mass energy of
one solar mass. Second, the variability timescale (few ms) observed
implies the energy deposition takes place in a small region of space
(radius of $c \times 1$ ms $\approx 30$ km). Third, the inferred rate
of GRB occurrence, about 4 per day in the Universe above current
detection thresholds, and the lack of burst repetition
\citepeg{hmp+98} suggest that GRB events are rare \citep[$\sim
10^{-7}$ yr$^{-1}$ Galaxy$^{-1}$;][]{feh93,wbbn98} and
catastrophically destroy the individual progenitors.

The progenitor models which most naturally explain these observables
in the GRB phenomena fall in to two broad classes---the coalescence of
binary compact stellar remnants and the explosion of a massive star
(``collapsar'').  An active galactic nucleus (AGN) origin is another
possibility, however the variability timescale still requires the
energy source to be stellar-mass objects \citep{car92,cw99}.  We
briefly summarize the popular progenitor models and refer the reader
to \citet{fwh99} for a more in-depth review. In both the collapsar and
the merging remnant class of progenitors a spinning BH is formed.  The
debris, either from the stellar core of the collapsar or a tidally
disrupted neutron star, forms a temporary accretion disk (or
``torus'') which then falls into the BH releasing a fraction of
gravitational potential of the matter. In this general picture
\citep[see][for a review]{ree99}, the lifetime of the accretion disk
accounts for the duration of the GRB and the light-crossing time of
the BH accounts for the variability timescale.  The GRB is powered by
the energy extracted either from the spin energy of the hole or the
from the gravitational energy of the in-falling matter.

The coalescing compact binary class \citep{pac86,goo86,eic+89} was
favored before the first redshift determination because the existence
of coalescence events of a double neutron star binaries (NS--NS) was
assured; at least a few NS--NS systems in our Galaxy (e.g., PSR
1913+16, PSR 1534+12) will merge in a Hubble time thanks to the
gravitational radiation of the binary orbital angular momentum
\citepcf{tay94}.  Further, the best estimate of the rate of NS--NS
coalescence in the Universe \citepeg{phi91,npp92} was comparable to an
estimate of the GRB rate \citep{feh93,wbbn98}.  Recently, stellar
evolution models have suggested that black hole--neutron star binaries
(BH--NS) may be formed at rates comparable to or even higher than
NS--NS binaries \citepeg{bb98}, though no such systems have been
observed to-date.  There are other merging remnant binaries which may
form GRBs, notably merging black hole--white dwarf (BH--WD) binaries
\citep{fw98} and black hole--helium star binaries (BH--He)
\citepcf{fwh99}.

The collapsar class is comprised of a rotating massive star, either
isolated or in a binary system, whose iron core subsequently collapses
directly to form a black hole \citep{woo93}.  To avoid baryon loading
the progenitor star should have lost most, if not all, of its extended
gas envelope of hydrogen by the time of collapse.  The progenitors of
collapsars---likely Wolf-Rayet stars---are then closely related to the
progenitors of hydrogen-deficient supernova, namely type Ib/Ic
supernovae \citepcf{mw99}. Perhaps one distinguishing difference is
that high angular momentum is necessary in collapsars. High angular
momentum centrifugally supports a transient torus around the BH,
creating a natural timescale for mass-energy injection.  The
efficiency of energy conversion is also helped around a spinning BH.
Further, angular momentum creates a natural rotation axis along which
large density gradients allow for the expanding blastwave to reach
relativistic speeds.

How can this large variety of viable GRB progenitors be distinguished?
Direct associations with other known astrophysical entities is
possible.  For massive stars, the energy release from the collapse of
the core of the star, just as in supernovae, is sufficient to explode
the star itself.  This may result in a supernova-like explosion at
essentially the same time as a GRB.  The first apparent evidence of
such a supernova associated with a cosmological GRB came with the
discovery of a delayed bright red bump in the afterglow light curve of
GRB 980326 \citep{bkd+99}.  The authors interpreted the phenomena as
due to the light-curve peak of a supernova at redshift $z \sim
1$. Later, \citet{rei99} and \citet{gtv+00} found similar such red
bump in the afterglow of GRB 970228.  Merging remnant progenitors
models (e.g., BH--NS, NS--NS systems) have difficulty producing these
features in a light curve on such long timescales and so the supernova
interpretation, if true, would be one of the strongest direct clues
that GRBs come from massive star explosions. However, the supernova
story is by no means complete.  For instance, in only one other GRB
(000911) has marginal ($\sim$2 $\sigma$) evidence of a SN signature
been found \citep{laz+01}; further, many GRBs do not appear to show
any evidence of SNe signatures \citepeg{hhc+00}. Even the
``supernova'' observations themselves find plausible alternative
explanations (such as dust echoes) that do not strictly require a
massive star explosion \citep{eb00,rei01,wd00}. We note, however, that
all other plausible explanations of the observed late-time bumps
require high-density environments found most readily in star forming
regions.

\citet{cl99b} emphasize that if a GRB comes from a massive star, then
the explosion does not take place in a constant density medium, but in
a medium enriched by constant mass loss from the stellar winds.  One
would expect to see signatures of this wind-stratified medium in the
afterglow \citep[e.g., bright sub-millimeter emission at early times,
increasing ``cooling frequency'' with time; see][]{pk00b,kbb+00}.
However, afterglow observations have been inconclusive \citep{kbb+00}
with no unambiguous inference of GRB in such a medium.

Recent work has begun to focus on the immediate environments of GRBs
as a means toward divining the nature of the progenitors.  This has
come primarily from detections of line features in now five GRB
afterglows \citep[e.g., 970508 and 970828][]{pcf+99,yno+99}. The most
recent and convincing detection so far comes from observations of the
afterglow of GRB 991216 \citep{pir01}.  Individually, the
observational significance of the line detections are marginal but on
the whole there appears to be a good case for line emission features
in the afterglow of some GRBs. If so, the inescapable conclusion is
that there must exist dense matter in the vicinity of the explosion
\citepeg{wmkr00,vpps99,lcg99a}, a seeming discordance with the
expectations of NS--NS merger models.

We emphasize that even the connection of GRBs to stellar-mass
progenitors has yet to be established.  The most compelling arguments
we have outlined (e.g.~temporal variability) rely on theoretical
interpretations of the GRB phenomena.  Further, direct observational
results (SNe signatures and transient Fe-line emission) are not yet
conclusive.

In this paper we examine the observed locations of GRBs with respect
to galaxies.  We find an unambiguous correlation of GRB locations with
the UV light of their hosts, providing strong indirect evidence for
the connection of GRBs to stellar-mass progenitors. Beyond this
finding, we aim to use the location of GRBs to distinguish between
stellar-mass progenitor models. In \S \ref{sec:location} we review the
expectations of GRB locations from each progenitor model.  Then in
\S\S \ref{sec:offdata}--\ref{sec:astlevels} we discuss the
instruments, techniques, and expected uncertainties involved in
constructing a sample of GRB locations about their host galaxies.  In
\S \ref{sec:indivoff} we comment on the data reductions specific to
each GRB in our sample.  The observed distribution is shown and
discussed in \S \ref{sec:offdist} and then statistically compared with
the expected offset distribution of leading progenitor models (\S
\ref{sec:compare}). Last, in \S \ref{sec:offsum} we summarize and
discuss our findings.

\section{Location of GRBs as a Clue to their Origin}
\label{sec:location}

Before the detailed modeling of light curves were used to constrain
the nature of supernovae progenitors, the location of supernovae in
and around galaxies provided important clues to the nature of the
progenitors \citepeg{rea53,jl63}.  For instance, only Type Ia
supernovae have been found in elliptical galaxies naturally leading to
the idea that the progenitor population can be quite old whereas the
progenitors of Type II and Type Ibc are likely to be closely related
to recent star formation \citep[cf.][for review]{vand92}.  Further, in
late-type galaxies, Type Ibc and Type II supernovae appear to be
systematically closer to HII star forming regions than Type Ia
supernovae \citepeg{btf94}.  This is taken as strong evidence that
the progenitors of Type Ibc and Type II SNe are massive stars
\citepcf{fil97}.

Since the most massive stars explode soon ($\ale 10^{7}$ yr) after
zero-age main sequence (ZAMS), we expect GRBs from collapsars to be
observed in galaxies undergoing vigorous star formation ({\it i.e.},
late-type, irregular, and starburst galaxies).  Merging neutron stars
on the other hand require a median time to merge of \hbox{$\sim
2$--$10 \times 10^{8}$~yr} since ZAMS
\citepeg{phi91,npp92,pzs96,bsp99}.  The instantaneous rate of GRBs
from binary mergers, then, is more a function of the integrated (as
opposed to instantaneous) star formation rate in its parent galaxy.
So if GRBs arise from the death of massive stars we do not expect
early-type ({\it i.e.}, elliptical and S0) host galaxies, whereas GRBs
from merging remnants could occur in such galaxies.  In principle, due
to the significant time from ZAMS to the mergers of NS--NS and BH--NS
binaries, such merging remnants should produce GRBs at preferentially
{\it lower} redshift than collapsars and promptly bursting binaries
(BH--He). In practice, though, distinguishing the GRB($z$) rate from
the SFR($z$) rate is extremely difficult without tens if not hundreds
more GRB redshift measures \citepeg{bsp99}.

More importantly, independent of galaxy type, the locations of GRBs
within (or outside) galaxies provide a powerful clue towards
distinguishing the progenitor scenarios. Massive stellar explosions
occur very near their birth-site, likely in active HII star-forming
regions, since the time since ZAMS is so small.  BH--He binaries will
merge quickly and so are also expected to be located near star-forming
regions \citep{fwh99}.  In stark constant, NS--NS and NS--BH binaries
merge far from their birthsite. These stellar remnant progenitors will
merge after at least one of the binary members has undergone a
supernova.  Each supernova is thought to impart a substantial ``kick''
on the resulting neutron star \citep[cf.][]{hp97}; for those binary
systems which survive both supernovae explosions, the center--of--mass
of the remnant binary itself will receive a velocity boost on the
order of a few hundred km/s \citepeg{bp95}.  That is, NS--NS or NS--BH
binaries will be ejected from their birthsite.  The gradual angular
momentum loss in the binary due to gravitational radiation causes the
binary to coalesce (or ``merge'') which then leads to a GRB.  The
exact time until merger ($\sim 10^{6}$--$10^{9}$ yr) depends on the
masses of the remnants and binary orbit parameters.  Population
synthesis models have all shown that roughly one third to one half of
NS--NS and BH--NS binary mergers will occur beyond 10 kpc in
projection from the centers of their hosts \citep{bsp99,fwh99}.  The
exact distribution of merger sites depends sensitively on the
gravitational potential of the host and the (radial) distribution of
massive star birth sites.

How have locations of GRBs within (or outside) galaxies impacted our
understanding of the progenitors of GRBs thus far?  As mentioned
above, the first accurate localization \citep{vgg+97} of a GRB by way
of an optical transient afterglow revealed GRB 970228 to be spatially
coincident with a faint galaxy \citep{slp+97,fpt+99,bdk01}.  Though
the nearby galaxy was faint, \citet{vgg+97} estimated the {\it a
posteriori} probability of a random location on the sky falling so
close to a galaxy by chance to be low.  As such, the galaxy was
identified as the host of GRB 970228.  \citet{slp+97} further noted
that the OT appeared offset from the center of the galaxy thereby
calling into question an active galactic nucleus (AGN) origin.  Soon
thereafter \citet{bdkf98} found, and then \citet{fp98} confirmed, that
GRB 970508 was localized very near the center of a dwarf galaxy. Given
that underluminous dwarf galaxies have a weaker gravitational
potential with which to bind merging remnant binaries, both
\citet{pac98b} and \citet{bdkf98} noted that the excellent spatial
coincidence of the GRB with its putative host found an easier
explanation with a massive star progenitor rather than NS--NS
binaries.

Once the afterglow fades, one could study in detail its environment
(analogous to low-redshift supernovae). Unfortunately, however, the
current instrumentation available for GRB observations cannot pinpoint
or resolve individual GRB environments on the scale of tens of parsecs
unless the GRB occurs a low redshift ($z \ale 0.2$) and the transient
afterglow is well-localized.  At higher redshifts (as all GRBs
localized to-date), only the very largest scales of galactic structure
can be resolved (e.g.~spiral arms) even by HST.  Therefore, the
locations of most individual GRBs do not yield much insight into the
nature of the progenitors. Instead, the observed {\it distribution} of
GRBs in and around galaxies must be studied as a whole and then
compared with the expectations of the various progenitor models.  This
is the aim of the present study. As we will demonstrate, while
not all GRBs are well-localized, the overall distribution of GRB
offsets proves to be a robust clue to the nature of the progenitors.

In this paper we present a sample of GRB offset measurements that
represents the most comprehensive and uniform set compiled
to-date. Every GRB location and host galaxy image has been re-analysed
using the most uniform data available.  The compilation is complete
with well-studied GRBs until May 2000. Throughout this paper we assume
a flat $\Lambda$--cosmology \citepeg{dab+00} with $H_0 = 65$ km
s$^{-1}$ Mpc$^{-1}$, $\Omega_M = 0.3$, and $\Lambda_0 = 0.7$.

\section{The Data: Selection and Reduction}
\label{sec:offdata}

The primary goal of this paper is to measure the offsets of GRBs from
their hosts where the necessary data are available.  Ideally this
could be accomplished using a dataset of early-time afterglow and
late-time host imaging observed using the same instrument under
similar observing conditions.  The natural instrument of choice is HST
given its exquisite angular resolution and astrometric stability.
Though while most hosts have been observed with HST at late-times,
there are only a handful of early-time HST detections of GRB
afterglow.  On the other hand, early ground-based images of GRB
afterglows are copious but late-time seeing-limited images of the
hosts give an incomplete view of the host as compared to an HST image
of the same field.  Moreover, ground-based imaging is inherently
heterogeneous, taken with different instruments, at different
signal-to-noise levels, and through a variety observing  of conditions;
this generally leads to poorer astrometric accuracy.  Bearing these
imperfections in mind we have compiled a dataset of images that we
believe are best suited to find offsets of GRBs from their hosts.

A listing of the dataset compilation is given in Table
\ref{tab:offsets}.  We include every GRB (up to and including GRB
000418) with an accurate radio or optical location and a deep
late-time optical image. There is a hierarchy of preference of imaging
conditions and instruments which yield the most accurate offsets; we
describe the specifics and expected accuracies of the astrometric
technique in \S \ref{sec:astlevels}.

\subsection{Dataset selection based on expected astrometric accuracy}
\label{sec:dataselect}

We group the datasets into 5 different levels ordered by decreasing
astrometric accuracy.  Levels 1--4 each utilized differential
astrometry and level 5 utilizes absolute astrometry relative to the
International Coordinate Reference System (ICRS). Specifics of the
individual offset measurements are given in \S
\ref{sec:astlevels}. The ideal dataset for offset determination is a
single HST image where both the transient and the host are
well-localized (hereafter ``self-HST''); so far, only GRB 970228, GRB
990123 and possibly GRB 991216 fall in this category. The next most
accurate offset is obtained where both the early- and late-time images
are from HST taken at comparable depth with the same filter (hereafter
``HST$\rightarrow$HST''). In addition to the centering errors of the
OT and host, such a set inherits the uncertainty in registering the
two epochs (e.g., GRB 970508).  Next, an early deep image from
ground-based (GB) Keck, Palomar 200-inch (P200), or the Very Large
Telescope (VLT) in which the OT dominates is paired with a late-time
image from HST (e.g.,~GRB 971214, GRB 980703, GRB 991216, GRB 000418;
``GB$\rightarrow$HST'').  Though in the majority of these cases most
of the objects detected in the HST image are also detected in the Keck
image (affording great redundancy in the astrometric mapping
solution), object centering of ground-based data is hampered by
atmospheric seeing.  The next most accurate localizations use
ground-based to ground-based imaging to compute offsets
(``GB$\rightarrow$GB'').  Last, radio localizations compared with
optical imaging (``RADIO$\rightarrow$OPT'') provide the least accurate
offset determinations. This is due primarily to the current difficulty
of mapping an optical image onto an absolute coordinate system (see \S
\ref{sec:vlahst}).

\subsection{Imaging Reductions}

\subsubsection{Reductions of HST Imaging}

Most of the HST images of GRB afterglow and host were acquired using
the {\it Space Telescope Imaging Spectrograph}
\citep[STIS;][]{kwb+98}.  STIS imaging under-samples the angular
diffraction limit of the telescope and therefore individual HST images
essentially do not contain the full astrometric information
possible. To produce a final image that is closer to the diffraction
limit, interpixel dithering between multiple exposures is often
employed. The image reconstruction technique, which also facilitates
removal of cosmic-rays and corrects for the known optical field
distortion, is called ``drizzling'' and is described in detail in
\cite{fh97}.  We use this technique, as implemented using the
IRAF\footnotemark\footnotetext{IRAF is distributed by the National
Optical Astronomy Observatories, which are operated by the Association
of Universities for Research in Astronomy, Inc., under cooperative
agreement with the National Science Foundation.} package DITHER and
DITHERII, to produce our final HST images.

We retrieved and reduced every public STIS dataset of GRB imaging from
the HST archive\footnotemark\footnotetext{{\tt
http://archive.stsci.edu}} and processed the so-called ``On--the--Fly
Calibration'' images to produce a final drizzled image.  These images
are reduced through the standard HST pipeline for bias subtraction,
flat-fielding, and illumination corrections using the best calibration
data available at the time of archive retrieval.  The archive name of
the last image and the start time of each HST epoch are given in
columns 2 and 3 of Table \ref{tab:offset-log}.

Some HST GRB imaging has been taken using the STIS/Longpass filter
(F28x50LP) which, based on its red effective wavelength (central
wavelength $\lambda_c \approx 7100$\AA), would make for a good
comparison with ground-based $R$-band imaging.  However, the Longpass
filter truncates the full STIS field of view to about 40\% and
therefore systematically contains fewer objects to tie astrometrically
to ground-based images.  Therefore, all of the HST imaging reported
herein were taken in (unfiltered) STIS/Clear (CCD50) mode.  Unlike the
Longpass filter, the spectral response of the Clear mode is rather
broad (2000--10000 \AA).  We use the known optical distortion
coefficients appropriate to the wavelength of peak sensitivity
$\lambda \approx 5850$ \AA\ of this observing mode to produce final images
which are essentially linear in angular displacement versus
instrumental pixel location.

The original plate scale of most STIS imaging is 0\arcsec.05077 $\pm$
0.00007 pixel$^{-1}$ \citep{mb97}, though there is a possibility that
thermal expansion of the instrument could change this scale by a small
amount (see Appendix \ref{sec:errors}). The pixel scale of all our
final reduced HST images is half the original scale, {\it i.e.},
0\arcsec.02539 pixel$^{-1}$.

\subsubsection{Reductions of Ground-based Imaging}

Ground-based images are all reduced using standard practice for bias
subtraction, flat-fielding, and in the case of $I$-band imaging, fringe
correction.  In constructing a final image we compute the instrumental
shift of dithered exposures relative to a fiducial exposure and co-add
the exposures after applying the appropriate shift to align each
image.  All images are visually inspected for cosmic-ray contamination
of the transient, host or astrometric tie stars.  Pixels contaminated
by cosmic rays are masked and not used in the production of the final
image.

\section{Astrometric Reductions and issues related to dataset levels}
\label{sec:astlevels}

Here we provide a description of the astrometric reduction techniques
for both our ground-based and the HST images, and issues related to the
five levels of astrometry summarized in \S \ref{sec:dataselect}. A
discussion of the imaging reductions and astrometry for the individual
cases is given in section \S \ref{sec:indivoff}.

\subsection{Level 1: self-HST (differential)}

An ideal image is one where the optical transient and the host galaxy
are visible in the same imaging epoch with HST.  This typically
implies that the host galaxy is large enough in extent to be
well-resolved despite the brilliance of the nearby OT. Of course, a
later image of the host is always helpful to confirm that the putative
afterglow point source does indeed fade. In this case (as with GRB
970228 and GRB 990123) the accuracy of offset determination is limited
mostly by the centroiding errors of the host ``center'' and optical
afterglow.  Uncertainties in the optical distortion corrections and
the resulting plate scale are typically sub-milliarcsecond in size
(see Appendix \ref{sec:errors}).

In principle we expect centering techniques to result in centroiding
errors ($\sigma_c$) on a point source with a signal-to-noise, SN, of
$\sigma_c \approx \phi/SN$ \citepcf{sto89}, where $\phi$ is the
instrumental full-width half maximum (FWHM) seeing of the final
image. Since $\phi$ is typically $\sim 75$ milliarcsecond (mas) we
expect $\sim$milliarcsecond offset accuracies with self-HST images.

\subsection{Level 2: HST$\rightarrow$HST (differential)}
\label{sec:hsthst}

Here, two separate HST epochs are used for the offset determination.
The first epoch is taken when the afterglow dominates the light and
the second when the host dominates.  In addition to the centroiding
errors, the astrometric accuracy of this level is limited by
uncertainty in the registration between the two images.

In general when two images are involved (here and all subsequent
levels), we register the two images such that an instrumental position
in one image is mapped to the instrumental (or absolute position) in
the other image. The registration process is as follows.  We determine
the noise characteristics of both of the initial and final images
empirically, using an iterative sigma-clipping algorithm.  This noise
along with the gain and effective read noise of the CCD are used as
input to the IRAF/CENTER algorithm.  In addition we measure the radial
profile of several apparent compact sources in the image and use the
derived seeing FWHM ($\phi$) as further input to the optimal filtering
algorithm technique for centering \citep[OFILTER; see][]{dav87}.  For
faint stars we use the more stable GAUSSIAN algorithm. Both techniques
assume a Gaussian form of the point-spread function which, while not
strictly matched to the outer wings of the Keck or HST PSFs, appears
to reasonably approximate the PSF out to the FWHM of the images.

\begin{figure*}
\centerline{\psfig{file=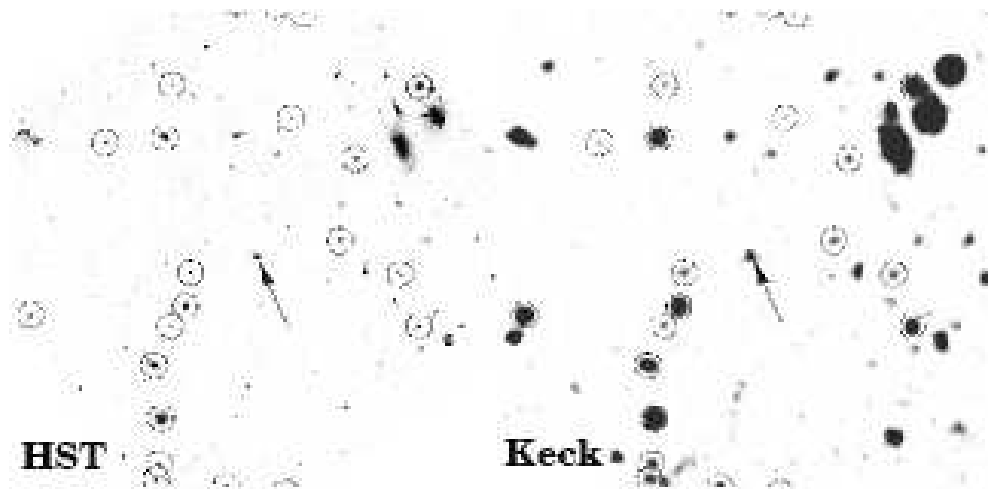,width=5.1in,angle=0}}
\caption[]{Example Keck $R$-band and HST/STIS Clear images of the field
of GRB 981226.  Twenty of the 25 astrometric tie objects are circled
in both images.  As with other Keck images used for astrometry in the
present study, most of the faint object detected in the HST images are
also detected (albeit with poorer resolution).  The optical transient
in the Keck image and the host galaxy in the HST image are in
center. The field is approximately 50\arcsec\ $\times$ 50\arcsec\ with
North up and East to the left. To view the full-resolution image, please download the
PDF version (or PS version) of the paper at
http://www.astro.caltech.edu/$\sim$jsb/Papers/offset.pdf (.ps.gz).}
\label{fig:examtie}
\end{figure*}

When computing the differential astrometric mappings between two
images (such as HST and Keck or Keck and the USNO-A2.0 catalog) we
use a list of objects common from both epochs, ``tie objects'', and
compute the astrometric mapping using the routine IRAF/GEOMAP.  The
polynomial order of the differential fitting we use depends on the
number of tie objects.  A minimum of 3 tie objects are required to
find the relative rotation, shift and scale of two images, which
leaves only one degree of freedom.  The situation is never this bad:
in fact, when comparing HST images and an earlier HST image (or deep
Keck image) we typically find 20--30 reasonable tie objects and
therefore we can solve for higher-order distortion terms. Figure
\ref{fig:examtie} shows an example Keck and HST field of GRB 981226
and the tie objects we use for the mapping. We always reject tie
objects that deviate by more than $3 \sigma$ from the initial
mapping. A full third-order 2-dimensional polynomial with cross-terms
requires 18 parameters which leaves, typically, $N \approx 30$ degrees
of freedom.  Assuming such a mapping adequately characterizes the
relative distortion, and it is reasonable to expect that mapping errors
will have an r.m.s.~error $\sigma \approx 30^{-1/2} \phi/\langle {SN}
\rangle$, where $\langle {SN} \rangle$ is the average
signal--to--noise of the tie objects.  For example, in drizzled HST
images $\phi \approx 75$ mas and $\langle {SN} \rangle \approx 20$ so
that we can expect differential mapping uncertainties at the 1 mas
level for HST$\rightarrow$HST mapping. Cross-correlation techniques,
such as IRAF/CROSSDRIZZLE, can in principle result in even better
mapping uncertainties, but, in light of recent work by \citet{ak99},
we are not confident that the HST CCD distortions can be reliably removed
at the sub-mas level.

\subsection{Level 3: GB$\rightarrow$HST (differential)}

This level of astrometry accounts for the majority of our dataset.  In
addition to inheriting the uncertainties of centroiding errors and
astrometric mapping errors described above, we must also consider the
effects of differential chromatic refraction (DCR) and optical image
distortion in ground based images.  In Appendix \ref{sec:errors} we
demonstrate that these effects should not dominate the offset
uncertainties. Following the argument above (\S \ref{sec:hsthst}) the
astrometric mapping accuracies scale linearly with the seeing of
ground-based images which is typically a factor of 10--20 larger than the
effective seeing of the HST images.

An independent test of the accuracy of the transference of
differential astrometry from ground-based images to space-based
imaging is illustrated by the case of GRB 990123. In \citet{bod+99} we
registered a Palomar 60-inch (P60) image to a Keck image and thence to
an HST image. The overall statistical uncertainty introduced by this
process (see Appendix \ref{sec:osh-derive} for a derivation) is
$\sigma_r = 107$ mas (note that in the original paper we mistakenly
overstated this error as 180 mas uncertainty).  The position we
inferred was 90 mas from a bright point source in the HST image.  This
source was later seen to fade in subsequent HST imaging and so our
identification of the source as the afterglow from
P60$\rightarrow$Keck$\rightarrow$HST astrometry was vindicated. Since
the P60$\rightarrow$Keck differential mapping accounted for
approximately half of the error (due to optical field distortion and
unfavorable seeing in P60), we consider $\sim 100$ mas uncertainty in
Keck$\rightarrow$HST mapping as a reasonable upper limit to the
expected uncertainty from other cases.  In practice, we achieved
r.m.s.~accuracies of 40--70 mas (see Table \ref{tab:offsets}).

\subsection{Level 4: GB$\rightarrow$GB (differential)}

This level contains the same error contributions as in
GB$\rightarrow$HST level, but in general, the uncertainties are larger
since the centroiding uncertainties are large in both epochs.  The
offsets are computed in term of pixels in late-time image.  Just as
with the previous levels with HST, we assume an average plate scale to
convert the offset to units of arcseconds.  For LRIS \citep{occ+95}
and ESI \citep{em98} imaging we assume a plate scale of 0\arcsec.212
pixel$^{-1}$ and 0\arcsec.153 pixel$^{-1}$, respectively.  We have
found that these plate scales are stable over time to better than a
few percent; consequently, the errors introduced by any deviations
from these assumed plate scales are negligible.

\subsection{Level 5: RADIO$\rightarrow$OPT (absolute)}
\label{sec:vlahst}

Unfortunately, the accuracy of absolute offset determination is
(currently) hampered by systematics in astrometrically mapping deep
optical/infrared imaging to the ICRS.  Only bright stars ($V \ale 9$
mag) have absolute localizations measured on the milliarcsecond level
thanks to astrometric satellite missions such as Hipparcos.  The
density of Hipparcos stars is a few per square degree so the
probability of having at least two such stars on a typical CCD frame
is low.  Instead, optical astrometric mapping to the ICRS currently
utilizes ICRS positions of stars from the USNO-A2.0 Catalog,
determined from scanned photographic plates \citep{mon98}.  Even if
all statistical errors of positions are suppressed, an astrometric
plate solution can do no better than inherit a systematic 1-$\sigma$
uncertainty of 250 mas in the absolute position of any object on the
sky \citep[$\sigma_\alpha = 0\arcsec.18$ and $\sigma_\delta =
0\arcsec.17$;][]{deu99}. By contrast, very-long baseline array (VLBA)
positions of GRB radio afterglow have achieved sub-milliarcsecond
absolute positional uncertainties relative to the ICRS \citep{tbf+99}.
So, until optical systematics are beaten down and/or sensitivities at
radio wavelengths are greatly improved (so as to directly detect the
host galaxy at radio wavelengths), the absolute offset astrometry can
achieve 1-$\sigma$ accuracies no better than $\sim 300$ mas ($\approx
2.5$ kpc at $z=1$).  In fact, one GRB host (GRB 980703) has been
detected at radio wavelengths \citep{bkf01} with the subsequent offset
measurement accuracy improving by a factor of $\sim$3 over the optical
measurement determined herein (\S \ref{subsec:980703}).

There are three GRBs in our sample where absolute astrometry (level 5)
is employed. In computing the location of the optical transient
relative to the ICRS we typically use 20--40 \hbox{USNO A2.0}
astrometric tie stars in common with Keck or Palomar images.  We then
use IRAF/CCMAP to compute the mapping of instrumental position ($x$,
$y$) to the world coordinate system ($\alpha$, $\delta$).

\section{Individual Offsets and Hosts}
\label{sec:indivoff}

Below we highlight the specific reductions for each offset, the
results of which are summarized in Table \ref{tab:offsets}. In total
there are 21 bursts until May 2000 that have been reliably localized
at the arcsecond level and 1 burst with an uncertain association with
the nearby SN 1998bw (GRB 980425). In our analysis we do not include
\hbox{GRB 000210} \citep{scm+00} due to lack of late-time imaging
data.  Thereby, the present study includes 20 ``cosmological'' GRBs
plus the nearby SN 1998bw/GRB 980425.  Offset measurements should be
possible for the recent bursts \hbox{GRB 000630} \citep{hur+00},
\hbox{GRB 000911} \citep{hcm+00b}, \hbox{GRB 000926} \citep{hmg+00}
and GRB 010222 \citep{pir+01}.

To look for the hosts, we generally image each GRB field roughly a few
months to a year after the burst with Keck.  Typically these
observations reach a limiting magnitude of $R \approx 24$--26 mag
depending on the specifics of the observing conditions. If an object
is detected within $\sim 1$ arcsec from the afterglow position and has
a brightness significantly above the extrapolated afterglow flux at
the time of observation, this source is deemed the host (most GRB
hosts are readily identified in such imaging).  If no object is
detected, we endeavor to obtain significantly deeper images of the
field. Typically these faint host searches require 1--3 hours of Keck
(or VLT) imaging to reach limiting magnitudes at the $R \approx 27$
mag level.  If no object is detected at the location of the afterglow,
HST imaging is required and the host search is extended to limiting
magnitudes of $R \approx 28$--29 mag.  Only 3 hosts in our sample (GRB
990510, GRB 000301C, and GRB 980326) were first found using HST after
an exhaustive search from the ground.

Note that the assignment of a certain observed galaxy as the host of a
GRB is, to some extent, a subjective process.  We address the
question of whether our assignments are ``correct'' in \S
\ref{sec:angoffs} where we demonstrate on statistical grounds that at
most only a few assignments in the sample of 20 could be spurious. In
\S \ref{sec:angoffs} we also discuss how absorption/emission redshifts
help strengthen the physical connection of GRBs to their assigned
hosts.

Irrespective of whether individual assignments of hosts are correct,
we uniformly assign the nearest (in angular distance) detected galaxy
as the host.  In practice this means that the nearest object
(i.e.~galaxy) brighter than $R \simeq 25$--26 mag detected in Keck
imaging is assigned as the host.  In almost all cases, there is a
detected galaxy within $\sim$1 arcsecond of the transient position.
For the few cases where there is no object within $\sim$1 arcsecond,
deeper HST imaging {\it always} reveals a faint galaxy within $\sim$1
arcsecond. In most cases, the estimated probability that we have
assigned the ``wrong'' host is small (see \S \ref{sec:angoffs}).
After assigning the host, the center of host is then determined,
except in a few cases, as the centroid near the brightest component of
the host system. In a few cases where there is evidence for
significant low-surface brightness emission (e.g.~980519) or the host
center is ambiguous we assign the approximate geometric center as the
host center.

A summary of our offset results is presented in Table
\ref{tab:offsets}.  Since all our final images of the host galaxies
are rotated to the cardinal orientation before starting the
astrometric mapping process, these uncertainties are also directly
proportional to the uncertainties in $\alpha$ and $\delta$. It is
important to note, however, that the projected radial offset is a
positive-definite number and the probability distribution is not
Gaussian.  Thereby, the associated error ($\sigma_r$) in offset
measurements does not necessarily yield a 68\% confidence region for
the offset (see Appendix \ref{sec:osh-derive}) but is, clearly,
indicative of the precision of the offset measurement.

Once the offsets are determined from the final images, we then measure
the half-light radii of the host galaxies.  For extended hosts, the
value of the half-light radius may be obtained directly from aperature
curve-of-growth analysis.  However, for compact hosts, the
instrumental resolution systematically spreads the host flux over a
larger area and biases the measurement of the half-light radius to
larger values.  We attempt to correct for this effect (for all hosts,
not just compact hosts) by deconvolving the images with IRAF/SCLEAN
using an average STIS/Clear point spread function (PSF) derived from
10 stars in the final HST image of the GRB 990705 field (which were
obtained through low Galactic latitude). We then fit curve-of-growth
photometry about the host centers and determine the radius at which
half the detected light was within such radius. These values, along
with associated errors are presented in table \ref{tab:offnorm}.  We
tested that the PSFs derived at differing roll angles and epochs had
little impact upon the determined value of the half-light radius.

\subsection{GRB 970228}

The morphology and offset derivation have been discussed extensively
in \citet{bdk01} and we briefly summarize the results.  In the
HST/STIS image (Figure~\ref{fig:offset1}), the host appears to be
essentially a face-on late-type blue dwarf galaxy.  At the center is
an apparent nucleus manifested as a 6-$\sigma$ peak north of the
transient.  There is also an indication of arm-like structure
extending toward the transient.

This image represents the ideal for astrometric purposes (level 1):
both the transient and the host ``center'' are well-localizable in the
same high-resolution image. The transient appears outside the
half-light radius of the galaxy.

\subsection{GRB 970508}

The host is a compact, elongated and blue galaxy \citep{bdkf98} and is
likely undergoing a starburst phase.  The optical transient was
well-detected in the early time HST image \citep{pfb+98a} and the host
was well-detected (Figure~\ref{fig:offset1}) in the late-time image
\citep{fp98}.  We masked out a 2\arcsec\ $\times$ 2\arcsec\ region
around the OT/host and cross-correlated the two final images using the
IRAF/CROSSDRIZZLE routine.  We used the IRAF/SHIFTFIND routine on the
correlation image to find the systematic shift between the two epochs.
The resulting uncertainty in the shift was quite small, $\sigma =
0.013$, $0.011$ pix ($x$, $y$ direction).  We also found 37 compact objects
in common to both images and performed an astrometric mapping in the
usual manner. We find $\sigma = 0.344, 0.354$ pix in the ($x$,$y$
directions). We centered the OT and the host in the normal manner
using the IRAF/ELLIPSE task.

The resulting offset is given in Table \ref{tab:offsets} where we use
the more conservative astrometric mapping uncertainties from using the
tie objects, rather than the CROSSDRIZZLE routine.  As first noted in
\citet{bdkf98} (Keck imaging) and then in \citet{fp98} (HST imaging),
the OT was remarkably close to the apparent center of the host galaxy.
The P200$\rightarrow$Keck astrometry from \citet{bdkf98} produced an
r.m.s.~astrometric uncertainty of 121 mas, compared to an
r.m.s.~uncertainty of 11 mas from HST$\rightarrow$HST astrometry.  The
largest source of uncertainty from the HST$\rightarrow$HST is the
centroid position of the host galaxy.

\subsection{GRB 970828}

The host is identified as the middle galaxy in an apparent 3-component
system.  We discuss the host properties and the astrometry
(RADIO$\rightarrow$OPT) in more detail in \citet{d00}. The total
uncertainty in the radio to Keck tie is 506 mas ($\alpha$) and 376 mas
($\delta$).

\subsection{GRB 971214}

By all accounts, the host appears to be a typical $L_*$ galaxy at
redshift $z=3.42$.  The Keck$\rightarrow$HST astrometry is discussed
in detail in \citet{odk+98}.  The offset
uncertainty found was $\sigma_r$ = 70 mas. The GRB appears located to
the east of the host galaxy center, but consistent with the east-west
extension of the host (see Figure \ref{fig:offset1}).

\subsection{GRB 980326}
\label{subsec:980326}

No spectroscopic redshift for this burst was found. However, based on
the light-curve and the SN hypothesis, the presumed redshift is
$z\sim$1 \citep[see][]{bkd+99}. \citet{bkd+99} reported that no galaxy
was found at the position of the optical transient down to a
3-$\sigma$ limiting magnitude of $R \approx 27.3$ mag.  Given the
close spatial connection of other GRBs with galaxies, we posited that
a deeper integration would reveal a nearby host.  Indeed,
\citet{fvn01} recently reported the detection with HST/STIS imaging of
a very faint ($V = 29.25 \pm 0.25$ mag) galaxy within 25 mas of the OT
position.

For astrometry we used an $R$-band image from 27 April 1998 when the
OT was bright and found the position of the OT on our deep $R$-band
image from 18 December 1998.  In this deep $R$-band image we found 34
objects in common with the HST/STIS drizzled image.  We confirm the
presence of this faint and compact source near the OT position though
our astrometry places the OT at a distance of 130 mas ($\sigma_r = 68$
mas).  

The galaxy and OT position are shown in Figure \ref{fig:offset1}.  The
low-level flux to the Southeast corner of the image is a remnant from
a diffraction spike of a nearby bright star. \citet{fvn01} find that
the putative host galaxy is detected at the 4.5-$\sigma$ level.
Adding to the notion that the source is not some chance superposition,
we note that the galaxy is the brightest object within 3 $\times$ 3
arcsec$^2$ of the GRB position.  There is also a possible detection of
a low-surface brightness galaxy $\sim$0.5 arcsec to the East of the
galaxy.

\subsection{GRB 980329}

The afterglow of GRB 980329 was first detected at radio wavelengths
\citep{tfk+98}.  Our best early time position was obtained using Keck
K-band image of the field observed by J.~Larkin and collaborators
\citep{lgk+98}.  We recently obtained deep R- imaging of the field
with Keck/ESI and detected the host galaxy at $R = 26.53 \pm 0.22$
mag.  We found the location of the afterglow relative to the host
using 13 stars in common to the early K-band and late R-band image.

As shown in Figure~\ref{fig:offset1}, the GRB is coincident with a
slightly extended faint galaxy.  Our determined angular offset
(cf.~Table \ref{tab:offsets}) of the GRB from this galaxy is
significantly closer to the putative host than the offset determined
by \citet{hth+00} in late-time HST imaging (our astrometric
uncertainties are also a factor of $\sim$ 9 smaller).  The difference
is possibly explained by noting that the \citet{hth+00} analysis used
the VLA radio position and just 3 USNO-A2.0 to tie the GRB position to
the HST image.

\subsection{GRB 980425}

The SN 1998bw was well-localized at radio wavelengths \citep{kfw+98}
with an astrometric position relative to the ICRS of 100 mas in each
coordinate.  Ideally we could calibrate the HST/STIS image to ICRS to
ascertain where the radio source lies. However, without
Hipparcos/Tycho astrometric sources or radio point sources in the STIS
field such absolute astrometric positioning is difficult.

Instead, we registered an early ground-based image to the STIS field
to determine the differential astrometry of the optical SN with
respect to its host.  Unfortunately, most early images were relatively
shallow exposures to avoid saturation of the bright SN and so
many of the point sources in the STIS field are undetected. The best
seeing and deepest exposure from ground-based imaging is from the
EMMI/ESO NTT 3.5 meter Telescope on 4.41 May 1998 \citep{gal99} where
the seeing was 0.9 arcsec FWHM.  We found 6 point sources which were
detected in both the STIS/CLEAR and the ESO NTT $I$-band image. The use
of $I$-band positions for image registration is justified since all 6
point sources are red in appearance and therefore unlikely to
introduce a systematic error in the relative positioning.  Since the
number of astrometric tie sources is low, we did not fit for
high-order distortions in the ESO image and instead we fit for the
relative scale in both the $x$ and $y$ directions, rotation, and shift (5
parameters for 12 data points).  We compute an r.m.s.~uncertainty of
40 mas and 32 mas in the $x$ and $y$ positions of the astrometric tie
sources. These transformation uncertainties dominate the error in the
positional uncertainty of the SN in the ESO NTT image and so we take
the transformation uncertainties as the uncertainty in the true
position of the supernova with respect to the STIS host image.

The astrometric mapping places the optical position of SN 1998bw
within an apparent star-forming region in the outer spiral arm of the
host 2.4 kpc in projection at $z=0.0088$ to the south-west of the
galactic nucleus.  Within the uncertainties of the astrometry the SN
is positionally coincident with a bright, blue knot within this
region, probably an HII region. This is consistent with the
independent astrometric solutions reported by \citet{fha+00}.

\subsection{GRB 980519}

The GRB afterglow was well-detected in our early-time image from the
Palomar 200-inch.  We found 150 objects in common to this image and
our intermediate-time Keck image. An astrometric registration between
the two epochs was performed using IRAF/GEOMAP.  Based on this
astrometry, \citet{bkdg+98} reported the OT to be astrometrically
consistent with a faint galaxy, the putative host. This is the second
faintest host galaxy (after GRB 990510; see below) observed to date
with $R = 26.1 \pm 0.3$.

We found 25 objects in common with the intermediate-time Keck image
and the HST/STIS image. These tie objects were used to further
propagate the OT position onto the HST frame.  Inspection of our final
HST image near the optical transient location reveals the presence of
low surface-brightness emission connecting the two bright elongated
structures.  Morphologically, the ``host'' appears to be tidally
interacting galaxies, although this interpretation is subjective.  The
GRB location is coincident with the dimmer elongated structure to the
north.  Using the approximate geometric center of the host we estimate
the center, albeit somewhat arbitrarily, as the faint knot south of
the GRB location and $\sim$0\arcsec .3 to the east of the brighter
elongated structure. The half-light radius of the system was also
measured from this point.  From this ``center'' we find the offset of
the GRB given in Table \ref{tab:offsets}.

\subsection{GRB 980613}

The morphology of the system surrounding the GRB is complex and
discussed in detail in \citet{dbk00}.  There we found the OT to be
within $\sim$3 arcsec of 5 apparent galaxies or galaxy fragments two
of which are very red ($R - K > 5$). In more recent HST imaging, the
OT appears nearly coincident with a compact high-surface brightness
feature, which we now identify as the host center.  Given the complex
morphology, we chose to isolate the feature in the determination of
the half-light radius by truncating the curve-of-growth analysis at
0.5 arcsec from the determined center.

\subsection{GRB 980703}
\label{subsec:980703}

The optical transient was well-detected in our early time image and,
based on the light curve and the late-time image, the light was not
contaminated by light from the host galaxy.
\citet{bkf01} recently found that the radio transient was very near
the center of the radio emission from the host. 

We found 23 objects in common to the Keck image and our final reduced
HST/STIS image and computed the geometric transformation.  The
r.m.s.~uncertainty of the OT position on the HST image was quite
small: 49 mas and 60 mas in the instrumental $x$ and $y$ coordinates,
respectively.  We determined the center of the host using
IRAF/ELLIPSE and IRAF/CENTER which gave consistent answers to 2
mas in each coordinate.  

Recently, \citet{bkf01} compared the VLBA position of the afterglow
with the position of the persistent radio emission from the
host. Since both measurements were referenced directly to the ICRS,
the offset determined was a factor of $\sim$3 times more accurate than
that found using the optical afterglow; the two offset measurements
are consistent within the errors.  In the interest of uniformity, we
will use the optical offset measurement in the following analysis.

\subsection{GRB 981226}

Unfortunately, no optical transient was found for this burst though a
radio transient was identified \citep{fkb+99}.  We rely on the
transformation between the USNO-A2.0 and the Keck image to place the
host galaxy position on the ICRS \citep[see][for further
details]{fkb+99}.  We then determined the location of the radio
transient in the HST frame using 25 compact sources common to both the
HST and Keck image. In Figure \ref{fig:examtie} we show as example the
tie objects in both the Keck and HST image.  The tie between the two
images is excellent with an r.m.s. uncertainty of 33 mas and 47 mas in
the instrumental $x$ and $y$ positions.  Clearly, the uncertainty in
the radio position on the Keck image dominates the overall location of
the GRB on the HST image.

The host appears to have a double nucleated morphology, perhaps
indicative of a merger or interacting system.  \citet{hta+00} noted,
by inspecting both the STIS Longpass and the STIS clear image, that
the north-eastern part of the galaxy appeared significantly bluer then
the south-western part. As expected from these colors the center of the
host as measured in our late-time $R$-band Keck lies near ($\sim 50$
mas) the centroid of the red (south-western) portion of the host.  We
assign the $R$-band centroid in Keck image as the center of the host.

\subsection{GRB 990123}

This GRB had an extremely bright prompt optical afterglow emission
which was found archivally in images from a robotic telescope, the
Robotic Optical Transient Search Experiment \citep{abb+99}.
We reported on the astrometric comparison of ground-based data with
HST imaging and found that the bright point source on the southern
edge of a complex morphological system was the afterglow
\citep{bod+99}.  Later HST imaging revealed that indeed this source
did fade \citepeg{ftp+99} as expected of GRB afterglow.

As seen in Figure \ref{fig:offset2}, the host galaxy is fairly
complex, with two bright elongated regions spaced by $\sim 0\arcsec.5$
which run approximately parallel to each other.  The appearance of
spatially curved emission to the west may be a tidal tail from the
merger of two separate systems or a pronounced spiral arm of the
brighter elongated region to the north.  We choose, again somewhat
subjectively, the peak of this brighter region as the center of the
system and find the astrometric position of the GRB directly from the
first HST epoch.

\subsection{GRB 990308}

An optical transient associated with GRB 990308 was found by
\citet{ssh+99}.  Though the transient was detected at only one epoch
\citep[3.3 hours after the GRB;][]{ssh+99} it was observed in three
band-passes, twice in $R$-band.  Later-time Keck imaging revealed no
obvious source at the location of the transient to $R$ = 25.7
suggesting that the source had faded by at least $\sim$7.5 mag in
$R$-band.

A deep HST exposure of the field was obtained by \citet{hft+00b} who
reported that the \citeauthor{ssh+99} position derived from the
USNO-A2.0 was consistent with two faint galaxies.

We found the offset by two means. First, we found an absolute
astrometric solution using 12 USNO-A2.0 stars in common with the
later-time Keck image.  The HST/STIS and the Keck $R$-band image were
then registered using 27 objects in common. Second, we found a
differential position by using early ground-based images kindly
provided by B.~Schaefer to tie the optical afterglow position directly
to the Keck (then to HST) image.  Both methods give consistent results
though the differential method is, as expected, more accurate.

Our astrometry places the OT position further East from the two faint
galaxies than the position derived by \citet{hft+00b}. At a distance of
0.73 arcsec to the North of our OT position there appears to be a
low-surface brightness galaxy near the detection limit of the STIS
image (see Figure~\ref{fig:offset2}), similar to the host of GRB 980519
(there is also, possibly, a very faint source 0.23 arcsec southwest of
the OT position, but the reality of its detection is questionable).
Due to the faintness and morphological nature of the source, a
detection confidence limit is difficult to quantify, but we are
reasonably convinced that the source is real.  At $V \sim 27$ mag, the
non-detection of this galaxy in previous imaging is consistent with
the current STIS detection. Since the angular extent of the galaxy
spans $\sim$25 drizzled STIS pixels ($\sim$0.63 arcsec), more
high-resolution HST imaging is not particularly useful for confirming
the detection of the galaxy. Instead, deeper ground-based imaging with
a large aperture telescope would be more useful.

\subsection{GRB 990506}

The Keck astrometric comparison to the radio position was given in
\citet{tbf+00}, with a statistical error of 250 mas. We transferred
this astrometric tie to the HST/STIS image using 8 compact sources
common to both the Keck and HST images of the field near GRB 990506.
The resulting uncertainty is negligible compared to the uncertainties
in the radio position on the Keck image.  As first reported in
\citet{tbf+00}, the GRB location appears consistent with a faint
compact galaxy.  \citet{htab+00} later reported that the galaxy
appears compact even in the STIS imaging.

\subsection{GRB 990510}
\label{subsec:990510}

This GRB is well-known for having exhibited the first clear evidence
of a jet manifested as an achromatic break in the light curve
\citepeg{hbf+99,sgk+99b}.  Recently, we discovered the host galaxy in
late-time HST/STIS imaging \citep{blo00} with $V = 28.5 \pm 0.5$.
Registration of the early epoch where the OT was bright reveals the OT
occurred 64 $\pm$ 9 mas west and 15 $\pm$ 12 mas north of the center
of the host galaxy.  This amounts to a significant displacement of 66
$\pm$ 9 mas or ~600 pc at a distance of $z = 1.62$ \citep{gvr+99}.
The galaxy is extended with a position angle PA = 80.5 $\pm$ 1.5
degree (east of north) and an ellipticity of about $\sim$0.5.

In retrospect, the host does appear to be marginally detected in the
July 1999 imaging as well as the later April 2000 image although at
the time, no galaxy was believed to have been detected
\citep{ffp+99}.

\subsection{GRB 990705}

\citet{mpp+00} discovered the infrared afterglow of GRB 990705
projected on the outskirts of the Large Magellanic Cloud.  At the
position of the afterglow, \citet{mpp+00} noted an extended galaxy
seen in ground-based $V$-band imaging; they identified this galaxy as
the host.  \citet{hah+00} reported on HST imaging of the field and
noted, thanks to the large size ($\sim 2$ arcsec) of the galaxy and
resolution afforded by HST, an apparent face-on spiral at the location
of the transient. We retrieved the public HST data and compared the
early images provided by N.~Masetti with our final reduced HST image.
Consistent with the position derived by \citet{hah+00}, we find that
the transient was situated on the outskirts of a spiral arm to the
west of the galaxy nucleus and just north of an apparent star-forming
region.

\subsection{GRB 990712}

This GRB is the lowest measured redshift of a ``cosmological'' GRB
with $z=0.4337$ \citep{hhc+00}.  Unfortunately, the astrometric
location of the GRB appears to be controversial, though there is no
question that the GRB occurred within the bright galaxy pictured in
Figure~\ref{fig:offset2}. \citet{hhc+00} found that the only source
consistent with a point source in the earlier HST image was the faint
region to the Northwest side of the galaxy and concluded that the
source was the optical transient.  However, \citet{fvh+00} found that
this source did not fade significantly.  Instead the Fruchter et
al.~analysis showed, by subtraction of two HST epochs, that a source
did fade near the bright region to the southeast.  While the fading
could be due to AGN activity instead of the presence of a GRB
afterglow, we adopt the conclusion of Fruchter et al.~for astrometry
and place conservative uncertainties on the location relative to the
center as 75 mas (3 pixels) in both $\alpha$ and $\delta$ for
1-$\sigma$ errors.  We did not conduct an independent analysis to
determine this GRB offset.

\subsection{GRB 991208}

In our early $K$-band image of the field, we detect the afterglow as
well as 7 suitable tie stars to our late ESI image.  The host galaxy
is visible in the ESI image and the subsequent offset was reported by us 
\citet{dbg+00}. An HST image was later obtained and reported by
\citet{fvsc00} confirming the presences of the host galaxy.

We reduced the public HST/STIS data on this burst and found the offset
in the usual manner by tying the OT position from Keck to the HST
frame.  The GRB afterglow position falls near a small, compact galaxy.
A fainter galaxy, to the Southeast, may also be related to the
GRB/host galaxy system (see Figure ~\ref{fig:offset2}).

\subsection{GRB 991216}

We used 9 compact objects in common to our early Keck image (seeing
FWHM = 0\arcsec.66) and the late-time HST/STIS imaging to locate the
transient.  As noted first by \citet{vff+00}, the OT is spatially
coincident with a faint, apparent point source in the HST/STIS image.
Our astrometric accuracy of $\sigma_r = 32$ mas of the OT position is
about 4 times better than that of \citet{vff+00}. Thanks to this we
can confidently state that the OT coincides with a point source on the
HST/STIS image. We believe this point source, as first suggested by
Vreeswijk et al., is the OT itself.

The ``location'' of the host galaxy is difficult to determine.  The OT
does appear to reside to the southwest of faint extended emission
(object ``N'' from Vreeswijk et al.~2000) but it is also located to
the northeast of a brighter extended component (object ``S'' from
Vreeswijk et al.~2000).  There appears to be a faint bridge of
emission connecting the two regions as well as the much larger region
to the west of the OT (see Figure~\ref{fig:offset2}).  In fact these
three regions may together comprise a large, low-surface brightness
system.  Again, somewhat arbitrarily, we take the center of the
``host'' to be the peak of object ``S''.

\subsection{GRB 000301C}

\citet{fmp+00}, in intermediate-time (April 2000) imaging of the field
of GRB 000301C, detected a faint unresolved source coincident with the
location of the GRB afterglow; the authors reckoned the source to be
the faded afterglow itself. In the most recent imaging on February
2001 the same group detected a somewhat fainter, compact object very
near the position of the transient.  Given the time ($\sim$ one year)
since the GRB the authors suggested that the afterglow should have
faded below the detection level and that therefore this object is the
host of GRB 000301C \citep{fv+01}

We confirm the detection of this source and measure the offset using
earlier time imaging from HST. Though no emission line redshift of
this source has been obtained given its proximity to the GRB it likely
resides at $z = 2.03$, inferred from absorption spectroscopy
\citep{sfg+01,cdd+00} of the OT.

In Figure \ref{fig:offset3} we present the late-time image from
HST/STIS.  A galaxy 2\arcsec.13 from the transient to the northwest is
detected at $R = 24.25 \pm 0.08$ mag and may be involved in possible
microlensing of the GRB afterglow \citepeg{gls01}

\subsection{GRB 000418}

We reported the detection of an optically bright component and an
infrared bright component at the location of GRB 000418
\citep{bdg+00}. \citet{mfm+00} later reported that HST/STIS imaging of
the field revealed that the OT location was 0\arcsec.08 $\pm$
0\arcsec.15 east of the center of the optically bright component, a
compact galaxy.

For our astrometry we used an early Keck $R$-band image and the late
HST/STIS image.  The astrometric uncertainty is improved over the
\citet{mfm+00} analysis by a factor of 2.4. Within errors, the OT is
consistent with the center of the compact host.

\section{The Observed Offset Distribution}
\label{sec:offdist}

\subsection{Angular Offset}
\label{sec:angoffs}

As seen in Table \ref{tab:offsets} and \S \ref{sec:indivoff}, there
are 20 GRBs for which we have a reliable offset measure from self-HST,
HST$\rightarrow$HST, HST$\rightarrow$GB, GB$\rightarrow$GB, or
RADIO$\rightarrow$OPT astrometric ties.  There are several
representations of this data worth exploring.  In Figure
\ref{fig:skyoff} we plot the angular distribution of GRBs about their
presumed host galaxy.  In this Figure and in the subsequent analysis
we exclude GRB 980425 because the association of this GRB with SN
1998bw is still controversial.  And more importantly (for the purposes
of this paper) the relation of GRB 980425 with the classical
``cosmological'' GRBs is unclear \citep{scm99} given that, if the
association proved true, the burst would have been under-luminous by a
factor of $\sim 10^{5}$ \citep{gvv+98,bkf+98}.

As can be seen from Table \ref{tab:offsets} and Figure
\ref{fig:skyoff}, well-localized GRBs appear on the sky close to
galaxies. The median projected offset of the 20 GRBs from their
putative host galaxies is 0.17 arcsecond---sufficiently small that
almost all of the identified galaxies must be genuine hosts (see
below). In detail, three of the bursts show no measurable offset from
the centroid of their compact hosts (970508, 980703, 000418) whereas
five bursts appear well displaced ($\age 0\arcsec.3$) from the center
of their host at a high level of significance.  Four additional bursts
detected via radio afterglows (GRB 970828, GRB 981226, GRB 990506) and
GRB 990308 (poor astrometry of the discovery image due to large pixels
and shallow depth) suffer from larger uncertainties
(r.m.s.~$\approx$0.3 arcsecond) but have plausible host galaxies.

As discussed in Appendix \ref{sec:errors}, GB$\rightarrow$GB or
GB$\rightarrow$HST astrometry could systematically suffer from the
effects of differential chromatic refraction, albeit on the
5--10 mas level. The HST$\rightarrow$HST measured offsets of GRB
970228, GRB 970508, GRB 990123, GRB 990510, GRB 990712, and GRB 000301C
are immune from DCR effects.  Since optical transients are, in
general, red in appearance and their hosts blue, DCR will
systematically appear to pull OTs away from their hosts in the
parallactic direction towards the horizon.  Comparing the observed
offsets directions parallactic at the time of each OT observation in
Table \ref{tab:offsets}, we find no systematic correlation thus
confirming that DCR does not appear to play a dominant role in
determining the differential offsets of OTs from their hosts.

On what basis can we be confident that the host assignment is correct
for a particular GRB? Stated more clearly in the negative is the
following question: "What is the probability of finding an unrelated
galaxy (or galaxies) within the localization error circle of the
afterglow (3-sigma) or, in the case where the localization error
circle is very small, whether a galaxy found close to a GRB
localization is an unrelated galaxy seen in projection?"  This
probability, assuming that the surface distribution of galaxies is
uniform and thus follows a Poisson distribution (i.e.~we ignore
clustering of galaxies) is
\begin{equation}
        P_{i,{\rm chance}} = 1 - \exp (-\eta_i).
\label{eq:thep}
\end{equation}
Here, 
\begin{equation}
        \eta_i = \pi r_i^2 \sigma (\leq m_i),
\nonumber
\end{equation}
is the expected number of galaxies in a circle with an effective
radius, $r_i$, and
\begin{equation}
  \sigma(\leq m_i) = \frac{1}{3600^2 \times 0.334 \log_e
10}\,10^{0.334\,\left(m_i - 22.963\right) + 4.320} {\rm ~~galaxy~arcsec}^{-2},
\nonumber
\end{equation}
is the mean surface density of galaxies brighter than R-band magnitude
of $m_i$, found using the results from \citet{hpm+97}. Since GRB are
observed through some Galactic extinction, the surface density of
galaxies at a given limiting flux is reduced; therefore, we use the
reddened host galaxy magnitude for $m_i$ (= col.~2 $-$ col.~3 of table
\ref{tab:offnorm}). 

There are few possible scenarios for determining $r_i$ at a given
magnitude limit. If the GRB is very well localized inside the
detectable light of a galaxy then $r_i \approx 2\, R_{\rm half}$, is a
reasonable estimate to the effective radius.  If the localization is
poor and there is a galaxy inside the uncertainty position, then $r_i
\approx 3\sigma_{R_0}$. If the localization is good, but the position
is outside the light of the nearest galaxy, then $r_i \approx
\sqrt{R_0^2 + 4\, R_{\rm half}^2}$. Therefore, we take $r_i = \max
[2\, R_{\rm half}, 3\sigma_{R_0}, \sqrt{R_0^2 + 4\, R_{\rm half}^2}]$
as a conservative estimate to the effective radius.  Here, the
quantity $R_0$ is the radial separation between the GRB and the
presumed host galaxy, $R_{\rm half}$ is the half-light radius, and
$\sigma_{R_0}$ is the associated r.m.s.~error (see Table
\ref{tab:offsets}).

If no ``obvious'' host is found (i.e.~$P_{\rm chance} \age 0.1$) then
we often seek deeper imaging observations, which will, in general,
decrease the estimated $r_i$ as more and more galaxies are
detected. However, the estimate for $\eta_i$ should remain reasonable
since the surface density of background galaxies continues to grow
larger with increasing depth. This is to say that there is little
penalty to pay in statistically relating sky positions to galaxies by
observing to fainter depths.

The values for $P_{i,{\rm chance}}$ are computed and presented in
Table \ref{tab:offnorm}. As expected, GRBs which fall very close to a
galaxy (e.g.~GRB 970508, GRB 980703, GRB 990712) are likely to be
related to that galaxy. Similarly, GRB localizations with poor
astrometric accuracy (e.g.~GRB 990308, 970828) yield larger
probabilities that the assigned galaxy is unrelated. 

In the past, most authors (including ourselves) did not endeavor to
produce a probability of chance association, instead opting to assume
that these assigned galaxies are indeed the hosts.  Nevertheless, we
believe these estimates are conservative; for instance, \citet{vgg+97}
estimated that $P_{\rm chance}$ (970228) = 0.0016 which is a value 5.8
times smaller than our estimate.  Again, we emphasize that the
estimated probabilities are constructed {\it a posteriori} so there is
no exact formula to the determine the true $P_{\rm chance}$.

The probability that {\bf all} supposed host galaxies in our sample
are random background galaxies is,
$$
P(n_{\rm chance} = m = {\rm all}) = \prod_{k=1}^m P_k,
$$
with $m = 20$ and $P_k$ found from equation \ref{eq:thep} for each GRB
$k$.  Not surprisingly, this number is extremely small, $P({\rm all})
= 2 \times 10^{-60}$, insuring that at least some host assignments must be
correct.  

The probability that {\bf all} galaxies are physically
associated (i.e.~that none are chance super-position of a random field
galaxies) is,
$$
P(n_{\rm chance} = 0) = \prod_{k=1}^m (1 - P_k) = 0.483.
$$
In general, the chance that $n$ assignments will be spurious out of a
sample of $m \ge n$ assignments is,
\begin{equation}
P(n_{\rm chance}) = \frac{1}{n_{\rm chance}{\rm !}} 
	\overbrace{\sum_i^m \sum_{j \ne i}^m \cdots\mathstrut}^{n_{\rm chance}}
	\left[\overbrace{P_i \times P_j \times \cdots\mathstrut}^{n_{\rm chance}}
	\prod_{k \not= i \not= j \not= \cdots}^m (1 - P_k)\right].
\end{equation}
$P(n_{\rm chance})$ reflects the probability that we have generated a
number $n_{\rm chance}$ of spurious host galaxy identifications.  For
our sample, we find that $P(1) = 0.395$, $P(2) = 0.106$, and $P(3) =
0.015$ and so the number of spurious identifications is likely to be
small, $\sim 1$--2. Indeed, if the two GRBs with the largest $P_{\rm
chance}$ are excluded (GRB 970828, GRB 990308) then $P(n_{\rm
chance}=0$) jumps to 0.76.  Thus we are confident that almost all of
our identifications are quite secure.

The certainty of our host assignments of the nearest galaxy to a GRB
finds added strength by using redshift information.  In {\it all}
cases where an absorption redshift is found in a GRB afterglow (GRB
970508, GRB 980613, GRB 990123, GRB 980703, GRB 990712, GRB 991216)
the highest redshift absorption system is observed to be at nearly
the same emission redshift of the nearest galaxy.  Therefore, with
these bursts, clearly the nearest galaxy cannot reside at a higher
redshift than the GRB.  The galaxy may simply be a foreground object
which gives rise both to nebular line emission and the absorption of
the afterglow originating from a higher redshift.  However, using the
observed number density evolution of absorbing systems, \citet{bsw+97}
calculated that statistically in $\age 80$\% of such absorption cases,
the GRB could reside no further than 1.25 times the absorption
redshift.  For example, if an emission/absorption system is found at
$z = 1.0$ then there is only a $\ale 20$\% chance that the GRB could
have occurred beyond redshift $z = 1.25$ without another absorption
system intervening.  Though this argument cannot prove that a given
GRB progenitor originated from the assigned host, the effect of
absorption/emission redshifts is to confine the possible GRB redshifts
to a shell in redshift-space, reducing the number of galaxies that
could possibly host the GRB, and increasing the chance that the host
assignment is correct. Therefore, given this argument and the
statistical formulation above, we proceed with the hypothesis that, as
a group, GRBs are indeed physically associated with galaxies assigned
as hosts.

\subsection{Physical Projection}
\label{sec:physproj}

Of the 20 GRBs with angular offsets five have no confirmed redshift
and the angular offset is thus without a physical scale.  These bursts
have hosts fainter than $R \approx 25$ mag and, given the distribution
of other GRB redshifts with these host magnitudes, it is reasonable
to suppose that the five bursts originated somewhere in the redshift
range $z =$ 0.5--5. It is interesting to note (with our assumed
cosmology) that despite a luminosity distance ratio of 37 between
these two redshifts, the angular scales are about the same:
$D_\theta(z=0.5)/D_\theta(z=5) \approx 1$.  In fact over this entire
redshift range, \hbox{6.6 kpc arcsec$^{-1}$ $< D_\theta(z) < 9.1$ kpc
arcsec$^{-1}$} which renders the conversion of angular displacement to
physical projection relatively insensitive to redshift.  For these
five bursts, then, we assign the median $D_\theta$ of the other bursts
with known redshifts so that \hbox{$D_\theta = 8.552$ kpc
arcsec$^{-1}$} (corresponding to a redshift of $z = 0.966$) and scale
the observed offset uncertainty by an additional 30\%.  Here, we use
the GRB redshifts (and, below, host magnitudes) compiled in the review
by \citet{kbb+00}. The resulting physical projected distribution is
depicted in Figure \ref{fig:skyoff} and given in Table 1. The median
projected physical offset of the 20 GRBs in the sample is 1.31 kpc or
1.10 kpc including only those 15 GRBs for which a redshift was
measured.  The minimum offset found is just 91 $\pm$ 90 pc from the
host center (GRB 970508).

\subsection{Host-Normalized Projected Offset}
\label{sec:hostnorm}

If GRBs were to arise from massive stars we would then expect that the
distribution of GRB offsets would follow the distribution of the light
of their hosts.  As can be seen in Figure \ref{fig:offset1},
qualitatively this appears to be the case since almost all
localizations fall on or near the detectable light of a galaxy. 

The next step in the analysis is to study the offsets but normalized
by the half-light radius of the host.  This step then allows us to
consider all the offsets in a uniform manner.  The half-light radius,
$R_{\rm half}$ is estimated directly from STIS images with
sufficiently high signal-to-noise ratio and in the remaining cases we
use the empirical half-light radius--magnitude relation of
\citet{owd+96}; we use the dereddened R-band magnitudes found in the
GRB host summaries from \citet{dfk+01a} and \citet{sfc+01}. Table
\ref{tab:offnorm} shows the angular offsets and the effective radius
used for scaling.  Where the empirical half-light radius--magnitude
relation is used, we assign an uncertainty of 30\% to $R_{\rm half}$

The median of the distribution of normalized offsets is 0.976 (Table
\ref{tab:offnorm}).  That this number is close to unity suggests a
strong correlation of GRB locations with the light of the host galaxy.
The same strong correlation can be graphically seen in Figure
\ref{fig:hostnorm} where we find that half of the galaxies lie inside
the half-light radius and the remaining outside the half-light radius.
We remark that the effective wavelength of the STIS band-pass and the
ground-based R band correspond to rest-frame UV and thus GRBs appear
to be traced quite faithfully by the UV light which mainly arises from
the youngest and thus massive stars. We will examine the distribution
in the context of massive star progenitors more closely in \S
\ref{sec:compare-collapsars}.

\subsection{Accounting for the uncertainties in the offset measurements}

A simple way to compare the normalized offsets to the expectations of
various progenitor models (see \S \ref{sec:compare}) is through the
histogram of the offsets. However, due to the small number of offsets,
the usual binned histogram is not very informative. In addition, the binned
histogram implicitly assumes that the observables can be represented
by $\delta$-functions and this is not appropriate for our case where
several offsets are comparable to the measurement uncertainty.

To this end we have developed a method to construct a probability
histogram (PH) that takes into account the errors on the
measurements. Simply put, we treat each measurement as a probability
distribution of offset (rather than a $\delta$-function) and create a
smooth histogram by summing over all GRB probability distributions.
Specifically, for each offset $i$ we create an individual PH
distribution function, $p_i(r)\,$d$r$, representing the probability of
observing a host-normalized offset $r$ for that burst. The integral of
$p_i(r)\,{\rm d}r$ is normalized to unity. The total PH is then
constructed as $p(r)\,$d$r$ = $\sum_i \, p_i(r)\,{\rm d}r$ and plotted
as a shaded region curve in Figure \ref{fig:offset-log}; see Appendix \ref{sec:osh-derive} for further details.

The total probability histogram $\int_0^{r} p(r)\,$d$r$ is depicted as
the solid smooth curve in Figure \ref{fig:offset-cum}.  There is, as
expected, a qualitative similarity between the cumulative total PH
distribution and the usual cumulative histogram distribution.
 
\section{Testing Progenitor Model Predictions}
\label{sec:compare}

Given the observed offset distribution, we are now in the position to
pose the question: which progenitor models are favored by the data?
Clearly, GRBs as a class do not appear to reside at the centers of
galaxies and so we can essentially rule out the possibility that {\it
all} GRBs localized to-date arise from nuclear activity.

\subsection{Delayed Merging Remnants Binaries (BH--NS and NS--NS)}

In general, the expected distributions of merging remnant binaries are
found using population synthesis models for high-mass binary evolution
to generate synthetic remnant binaries.  The production rate of such
binaries from other channels (such as three-body encounters in dense
stellar clusters) are assumed to be small relative to isolated binary
evolution. Due to gravitational energy loss, the binary members
eventually coalesce but may travel far from their birth-site before
doing so.  The locations of coalescence are determined by integrating
the synthetic binary orbits in galactic potential models.

\citet{bsp99}, \citet{fwh99} and \citet{bbz99} have simulated the
expected radial distribution of GRBs in this manner.  All three
studies essentially agree on the NS--NS differential offset
distributions as a function of host galaxy mass. (Note that the
displayed distance axis, when compared to the differential
distribution, is erroneously too large by a factor of ten in the {\it
cumulative} offset prediction plot in Figure~22 of Fryer et al.~1999.)
\citet{fwh99} suggest that the \citet{bsp99} synthesis may have
incorrectly predicted an over-abundance of compact binaries with small
merger ages, because the population synthesis did not include a
non-zero helium star radius; this is not the case although an
arithmetic error in our code may account for the discrepancy
(Sigurdsson, priv.~communication).

The formation scenarios of BH--NS binaries are less certain than that of
NS--NS binaries.  Both \citet{fwh99} and \citet{bbz00} suggest that
so-called ``hypercritical accretion'' \citep{bb98} dominates the
birthrate of BH--NS binaries.  Briefly, hypercritical accretion occurs
when the primary star evolves off the main sequence and explodes as a
supernova, leaving behind a neutron star.  Mass is rapidly accreted
from the secondary star (in red giant phase) during common envelope
evolution, causing the primary neutron star to collapse to a black
hole.  The secondary then undergoes a supernova explosion leaving
behind a NS. As in NS--NS binary formation, only some BH--NS systems
will remain bound after having received systemic velocity kicks from
two supernovae explosions.  

One important difference is that BH--NS binaries are in general more
massive (total system mass $M_{\rm tot} \approx 5 M_\odot$) than
NS--NS binaries ($M_{\rm tot} \approx 3 M_\odot$).  Furthermore, the
coalescence timescale after the second supernova is shorter that in
NS--NS binaries because of the BH mass.  Therefore, despite similar
evolutionary tracks BH--NS binaries could be retained more tightly to
host galaxies than NS--NS binaries \citep{bsp99,bbz00}. \citet{bbz00}
quantified this expected trend, showing that on average, BH--NS
binaries merge $\sim$few times closer to galaxies than NS--NS
binaries. Surprisingly, \citet{fwh99} found that BH--NS binaries
merged {\it further} from galaxies than NS--NS binaries, but this
result was not explained by \citet{fwh99}.  Nevertheless, just as with
NS--NS binaries, a substantial fraction of BH--NS binaries will escape
the potential well of the host galaxy and merge well-outside of the
host. For example, even in massive galaxies such as the Milky Way,
these studies show that roughly 25\% of mergers occur $> 100$ kpc from
the center of a host galaxy.

Before comparing in detail the predicted and observed distributions,
it is illustrative to note that the observed distribution appears
qualitatively inconsistent with the delayed merging remnant binaries.
All the population synthesis studies mentioned thus far find that at
approximately 50\% of merging remnants will occur outside of $\approx
10$ kpc when the mass of the host is less than or comparable to the
mass of the Milky Way. Comparing this expectation with
Figure~\ref{fig:skyoff}, where no bursts lie beyond 10 kpc from their
host, the simplistic Poisson probability that the observed
distribution is the same as the predicted distribution is no larger
than 2 $\times 10^{-3}$.

To provide a more quantitative comparison of the observed distribution
with the merging remnant expectation, we require a model of the
location probability of GRB mergers about their hosts.  These models,
which should in principle vary from host to host, have a complex
dependence on the population synthesis inputs, the location of star
formation within the galaxies and the dark-matter halo mass.

No dynamical or photometric mass of a GRB host has been reported
to-date. However, since GRB hosts are blue, starbursts
\citepeg{dfk+01a} it is not unreasonable to suspect that their masses
will lie in the range of 0.001 -- 0.1 $\times$ 10$^{11} M_\odot$
\citepeg{oab+01}. The most obvious exceptions to this are the hosts of
GRB 971214 and GRB 990705 which are likely to be near $L_*$.  The
observed median effective disk scale length of GRB hosts is $r_e =
1.1$ kpc though GRB hosts clearly show a diversity of sizes (Table
\ref{tab:offnorm}, col.~5).  This value of $r_e$ is also close to the
median effective scale radii found in the \citet{oab+01} study of
nearby compact blue galaxies.

To compare the observed and predicted distributions we use galactic
models a--e from \citet{bsp99} corresponding to hosts ranging in mass
from 0.009 -- 0.62 $\times 10^{11} M_\odot$ and disk scale radii
($r_e$) of 1 and 3 kpc.  Following the discussion above, we also
construct a new model ($a^*$) which we consider the most representative
of GRB hosts galaxies with $v_{\rm circ} = 100$ km s$^{-1}$, $r_{\rm
break} = 1$ kpc, and $r_e = 1.5$ kpc ($M_{\rm gal} = 9.2 \times 10^{9}
M_\odot$).

We project these predicted {\it radial} distribution models by
dividing each offset by a factor of $1.15$ since the projection of a
merger site on to the plane of the sky results in a smaller observed
distance to the host center than the radial distance. We determined
the projection factor of 1.15 by a Monte Carlo simulation projecting a
3-dimensional (3-D) distribution of offsets onto the sky. The median
projected offset is thus 87\% of the 3-D radial offset.  

The observed distribution is compared with the predicted distributions
and shown in Figure \ref{fig:rem-comp1}.  We summarize the results in
table \ref{tab:nsks}. Only model $d$ ($M = 6.3 \times 10^{10}
M_\odot$, $r_e = 3$ kpc) could be consistent with the data ($P_{\rm
KS} = 0.063$) but this galactic model has a larger disk and is
probably more massive than most GRB hosts.  Instead, for the ``best
bet'' model $a^*$, the one-sided Kolmogorov-Smirnov probability that
the observed sample derives from the same predicted distribution is
$P_{\rm KS} = 2.2\times 10^{-3}$, in agreement with our simplistic
calculation above; that is, the location of GRBs appears to be
inconsistent with the NS--NS and NS--BH hypothesis.

If GRBs do arise from systems which travel far from their birthsite,
then there is a subtle bias in determining the offset to the host.  If
the progenitors are ejected from the host by more than half the
distance between the host and the nearest (projected) galaxy, then the
transient position will appear unrelated to any galaxies (the wrong
host will be assigned, of course) but $P_{\rm chance}$ will always
appear high no matter how deep the host search is.  We try account for
this effect in our modeling (Appendix \ref{sec:robust}) by
synthetically replacing observed (small) offsets that are associated
with a high value of $P_{\rm chance}$ with new, generally larger,
offsets drawing from the expected distribution of offsets for a
particular galactic model.  This then biases the distribution of
$P_{KS}$ statistics toward {\it higher} values (by definition) but the
median values of $P_{KS}$ are largely unaffected (see table
\ref{tab:nsks}).

\subsection{Massive Stars (Collapsars) and Promptly Bursting
Binaries (BH--He)}
\label{sec:compare-collapsars}

As discussed, collapsars produce GRBs in star-forming regions, as will
BH--He binaries. The localization of GRB 990705 near a spiral arm is, of
course, tantalizing smaller-scale evidence of the GRB--star-formation
connection.  Ideally, the burst sites of individual GRBs could be
studied in detail with imaging and spectroscopy and should, if the
collapsar/promptly bursting binary origin is correct, reveal that the
burst sites are HII regions.  Unfortunately, the distances to GRBs
preclude a detailed examination of the specific burst sites at the
tens of parsec scale resolution (the typical size for a star-forming
region) with current instrumentation. Adaptive optics (AO) laser-guide
star imaging may prove quite useful in this regard as will IR imaging
with the {\it Next Generation Space Telescope}.

Weaker evidence for a star-formation connection exists in that no GRB
to date has been observed to be associated with an early-type galaxy
(morphologically or spectroscopically), though in practice it is often
difficult to discern galaxy type with the data at hand.  Indeed most
well-resolved hosts appear to be compact star forming blue galaxies,
spirals, or morphological irregulars.

Above we have demonstrated that GRBs follow the UV (restframe) light
of their host galaxies. However, the comparison has been primarily
mediated by a single parameter, the half-light radius and the median
normalized offset. We now take this comparison one step further. For
the GRB hosts with high signal-to-noise HST detections (e.g.~GRB
970508, GRB 971214, GRB 980703) our analysis shows that the surface
brightness is well-approximated by an exponential disk. We use this
finding as the point of departure for a simplifying assumption about
all GRB hosts: we assume an exponential disk profile such that the
surface brightness of the host galaxy scales linearly with the
galactocentric radius in the disk. We further assume that the star
formation rate of massive stars scales with the observed optical light
of the host; this is not an unreasonable assumption given that
HST/STIS imaging probes restframe UV light, an excellent tracer of
massive stars, at GRB redshifts.

Again, clearly not all host galaxies are disk-like
(Figure~\ref{fig:offset1}) so this assumption is not strictly valid in
all cases. If $r_e$ is the disk scale length, the half-light radius of
a disk galaxy is $R_{\rm half} = 1.67 \times r_e$, so that the
simplistic model of the number density of massive star-formation
regions in a galaxy is,
\begin{equation}
N(r)\, dr \propto r \exp(-1.67\, r)\, dr,
\label{eq:sfr}
\end{equation}
where $r = R/R_{\rm half}$.  In reality, the distribution of massive
star formation in even normal spirals is more complex, with a strong
peak of star formation in the nuclear region and troughs between
spiral arms \citepeg{rw86,bdd89}. We make an important assumption when
comparing the observed distribution with the star-formation disk
model: that each GRB occurs in the disk of its host (see discussion
below).  Dividing the observed offset by the apparent half-light
radius host essentially performs a crude de-projection.

We find the probability that the observed distribution could be
derived from the simplistic distribution of massive star regions
(eq.~\ref{eq:sfr}) is $P_{\rm KS} = 0.454$ (i.e.~the two distributions
are consistent).  In Appendix \ref{sec:robust} we show that these
results are robust even given the measurement uncertainties.  This
broad agreement between GRB positions and the UV light of their hosts
is remarkable in the sense that the model for massive star locations
is surely too simplistic; even in classic spiral galaxies (which most
GRB hosts are not) star-formation is a complex function of
galactocentric radius, with peaks in galactic centers and spiral arms.
Furthermore, surface brightness dimming with redshift causes galaxies
to appear more centrally peaked, resulting in a systematic
underestimate of $R_{\rm half}$.

\section{Discussion and Summary}
\label{sec:offsum}

We have determined the observed offset distribution of GRBs by
astrometrically comparing localizations of GRB afterglow with optical
images of the field surrounding each GRB. In all cases, the GRB
location appears ``obviously'' associated with a galaxy---either
because the position is superimposed atop a galaxy or very near ($\ale
1.2\arcsec$) a galaxy in an otherwise sparse field.  In fact,
irrespective of the validity of individual assignments of hosts, the
offset distribution may be considered a distribution of GRB positions
from the nearest respective galaxy at least as bright as $R \approx
28$ (note that in most cases the host galaxies are much brighter,
typically $R = 24$--26 mag). We find that at most a few of the 20 GRBs
could be unrelated physically to their assigned host and about a 50\%
chance that all GRBs are correctly assigned to their hosts (see \S
\ref{sec:angoffs}).

We then compare the distribution of GRB locations about their
respective hosts with the {\it predicted} radial offset distribution
of merging binary remnants. This comparison is complicated by an
unknown projection factor for each burst: if a GRB occurs near an
edge-on disk galaxy there exists no model-independent manner to
determine the true 3-D radial offset of the GRB from the center of the
host.  Indeed, in a few cases (e.g., GRB 980519, GRB 991216) even the
``center'' of the host is not well defined and we must estimate a
center visually.  In all other cases, we find the centers using a
luminosity-weighted centroid surrounding the central peak of the
putative host.

To compare the GRB offsets with those predicted by the NS--NS and
NS--BH binary models, we make a general assumption about the
projection factor and, to facilitate a comparison in physical units
(that is, offsets in kiloparsec rather than arcseconds), we assign an
angular diameter distance to the 5 hosts without a confirmed distance
(\S \ref{sec:physproj}).  We have shown that the conversion of an
angular offset to physical projection is relatively insensitive to the
actual redshift of the host. We estimate that the probability that the
observed GRB offset distribution is the same as the predicted
distribution of NS--NS and BH--NS binaries is $P \ale 2 \times
10^{-3}$.  Insofar as the observed distribution is representative (see
below) and the predicted distribution is accurate, our analysis
renders BH--NS and NS--NS progenitor scenarios unlikely for
long-duration GRBs.

Having cast doubt on the merging remnant hypothesis, we test whether
the offset distribution is consistent with the collapsar (or BH--He)
class.  Since massive stars (and promptly merging binaries) explode
where they are born, we have compared the observed GRB offset
distribution with a very simplistic model of massive star formation in
late-type galaxies: an exponential disk.  After normalizing each GRB
offset by their host half-light radius we compare the distribution
with a KS test and find good agreement: $P_{\rm KS} = 0.454$. We have
shown that these KS results, based on the assumption of
$\delta$-function offsets, are robust even after including the
uncertainties in the offset measurements.

Thus far we have neglected discussion of the observational biases that
have gone into the localizations of these 20 GRBs.  The usual problems
plaguing supernova detection, such as the brightness of the central
region of the host and dust obscuration, are not of issue for
detection of the {\it prompt} high-energy emission ({\it i.e.}, X-rays
and $\gamma$-rays) of GRBs since the high-energy photons penetrate
dust.  If the intrinsic luminosity of GRBs is only a function of the
inner-workings of the central engine (that is, GRBs arise from
internal shocks and not external shocks) then the luminosity of a GRB
is independent of ambient number density.  Therefore, prompt X-ray
localizations from BeppoSAX and $\gamma$-ray locations from the IPN
should not be a function of the global properties of GRB environment;
only intrinsic GRB properties such as duration and hardness will
affect the prompt detection probability of GRBs.

The luminosity of the afterglows is, however, surmised to be a function
of the ambient number density. Specifically, the afterglow luminosity
will scale as $\sqrt{n}$ where $n$ is the number density of hydrogen
atoms in the 1--10 pc region surrounding the GRB explosion site
\citep[cf.][]{mrw98b}.  While $n \approx$ 0.1--10 cm$^{-3}$ in the
interstellar medium, the ambient number density is probably $n \approx
10^{-4}$--10$^{-6}$ in the intergalactic medium.  Thus GRB afterglows
in the IGM may appear $\sim 10^{-3}$ times fainter than GRB afterglows
in the ISM (and even more faint compared to GRBs that occur in
star-forming regions where the number densities are higher than in the
ISM). If only a small fraction of GRBs localized promptly in X-rays
and studied well at optical and radio wavelengths were found as
afterglow, the ambient density bias may be cause for concern. However,
this is not the case. As of June 2001, 29 of 34 bursts localized by
prompt emission were later found as X-ray, optical, and/or radio
afterglow \citepcf{fkw+00}; that is, almost all GRBs have detectable
X-ray afterglow.  Therefore, {\it no more than about $10\%$ of GRBs
localized by BeppoSAX could have occurred in significantly lower density
environments} such as in the IGM; thus, we do not believe that our
claim against the delayed merging binaries is affected by this
bias. 

What about the non-detection of GRB afterglow at optical/radio
wavelengths?  Roughly half of GRBs promptly localized in the gamma-ray
or X-ray bands are not detected as optical or radio afterglow
\citep{fkw+99,lcg01}.  While many of these ``dark'' GRBs must be due
to observing conditions (lunar phase, weather, time since burst, etc.)
at least some fraction may be due to intrinsic extinction local to the
GRB.  If so, then these GRBs are likely to be centrally biased since
the optical column densities are strongest in star-forming regions and
giant molecular clouds.  Therefore, {\it any optically obscured GRBs
which do not make it in to our observed offset sample will be
preferentially located in the disk}. We do not therefore believe the
ambient density bias plays any significant role in causing GRBs to be
localized preferentially closer to galaxies; in fact, the opposite may
be true.

The good agreement between our simplistic model for the location of
massive stars and the observed distribution is one of the strongest
arguments yet for a collapsar (or promptly bursting binaries) origin
of long-duration GRBs. However, the concordance of the predicted and
observed distributions are necessary to prove the connection, although
not sufficient. 

We may now begin to relate the offsets to the individual host and GRB
properties. For instance, of the GRBs which lie in close proximity to
their host centers (GRB 970508, GRB 980703, and GRB 000418), there is
a striking similarity between their hosts---all appear compact and
blue with high-central surface brightness suggesting that these hosts
are nuclear starburst galaxies (none show spectroscopic evidence for
the presence of an AGN).  

In fact, the closeness of some GRBs to their host centers signifies
that our simplistic model for star-formation may require modification.
This is not unexpected since, in the Galaxy, star formation as a
function of Galactocentric radius does not follow a pure exponential
disk, but is vigorous near the center and is strongly peaked around $R
\sim 5$ kpc \citepcf{ken89}.  As more accurate offsets are amassed,
these subtle distinctions in the GRBs offset distribution may be
addressed.

\acknowledgments

The authors thank the staff of the W.~M.~Keck Foundation and the staff
at the Palomar Observatories for assistance.  We thank the anonymous
referee for very insightful comments; in particular, the referee
pointed out (and suggested the fix for) the bias in offset assignment
if GRBs are ejected far from their host galaxies. The referee also
pointed out that we did not establish that GRBs are preferentially
aligned with the major axes of their hosts (as we had claimed in an
earlier version of the paper). We applaud E.~Berger and D.~Reichart
for close reads of various drafts of the paper.  We also thank
M.~Davies for encouraging us to compare several NS--NS models to the
data rather than just one.  We acknowledge the members of the
Caltech-NRAO-CARA GRB collaboration and P.~van Dokkum, K.~Adelberger,
and R.~Simcoe for helpful discussions. We thank N.~Masetti for
allowing us access to early ground-based data on GRB 990705 and
B.~Schaefer for kindly providing {\it QUEST} images of the afterglow
associated with GRB 990308. This work was greatly enhanced by the use
of data taken as part of the {\it A Public Survey of the Host Galaxies
of Gamma-Ray Bursts} with HST (\#8640; S.~Holland, P.I.). JSB
gratefully acknowledges the fellowship and financial support from the
Fannie and John Hertz Foundation.  SGD acknowledges partial support
from the Bressler Foundation. SRK acknowledges support from NASA and
the NSF.  The authors wish to extend special thanks to those of
Hawaiian ancestry on whose sacred mountain we are privileged to be
guests. Without their generous hospitality, many of the observations
presented herein would not have been possible.


\appendix

\section{Potential sources of astrometric error}
\label{sec:errors}

\subsection{Differential Chromatic Refraction}

Ground-based imaging always suffers from differential chromatic
refraction (DCR) introduced by the atmosphere.  The magnitude of this
refraction depends strongly ($\propto 1/\lambda_{\rm eff}^2$) on the
effective wavelength ($\lambda_{\rm eff}$) of each object, the airmass
of the observation, and the air temperature and pressure.  With
increasing airmass, images are dispersed by the atmosphere and
systematically stretched in the parallactic direction in the sense
that bluer objects shift toward the zenith and redder objects shift
toward the horizon.  Other sources of refraction, such as turbulent
refraction \citepeg{lin80}, are statistical in nature and will only
serve to increase the uncertainty in our astrometric solution.

Here we show that DCR is, in theory, will not dominate our offset
determinations.  Since all of our early ground-based imaging were
conducted with airmass ($\sec(z)$) $\ale 1.6$ we take as an extreme
example an image with airmass $\sec (z) = 2$, where $z$ is the
observed zenith angle. It is instructive to determine the scale of
systematic offset shifts introduced when compared with either
late-time ground-based or HST imaging where refractive distortions are
negligible.  Following \citet{gt98}, the differential angular
distortion between two point sources at an apparent angular separation
along the zenith, $\Delta z$, may be broken into a color and a zenith
distance term.  Assuming nominal values for the altitude of the Keck
Telescopes on Mauna Kea, atmospheric temperature, humidity and
pressure, at an effective wavelength of the $R$-band filter,
$\lambda_{\rm eff} (R) = 6588$ \AA\ \citep{fig+96}, the zenith
distance term is 16 mas for an angular separation of 30 arcsec at an
airmass of $\sec (z) = 2$.  To first order, the zenith term is linear
in angular distance and so, in practice, even this small effect will
be accounted for as a first-order perturbation to the overall
rotation, translation, and scale mapping between a Keck and HST image.
In other words, we can safely neglect the zenith term contribution to
the DCR.

We now determine the color term contribution. Optical transients of
GRBs are in general, redder in appearance (apparent $V - R \approx
0.5$ mag) than their host galaxies (apparent $V - R \sim 0.2$ mag).
We assume the average astrometric tie object has $V-R = 0.4$ mag.  If
the OT is observed through an airmass of $\sec(z) = 2$ and then the
galaxy is observed at a later time through and airmass of, for
example, $\sec(z) = 1.2$, then DCR will induce a $\sim 30$ mas
centroid shift between the OT and the host galaxy if the two epochs
are observed in $B$-band \citep[cf.~Figure~2 of][]{aaa+99}.  In
$R$-band, the filter used in almost all of our ground-based imaging
for the present work, the DCR strength is about 20\% smaller than in
$B$-band because of the strong dependence of refraction on
wavelength. Therefore we can reasonably assume that DCR should only
{\it systematically} affect our astrometric precision at the 5---10
mas level.  Such an effect could, in principle, be detected as a
systematic offset in the direction of the parallactic angles of the
first epochs of GRB afterglow observations. In \S \ref{sec:angoffs} we
claim that no such systematic effect is present in our data. DCR could
of course induce a larger {\it statistical} scatter in the uncertainty
of an astrometric transformation between epochs since individual tie
objects are not, in general, the same color and each will thus
experience its own DCR centroid shift.

Bearing in mind that DCR is probably negligible we can minimize the
effects of DCR by choosing small fields and similar spectral responses
of the offset datasets.  The HST fields are naturally small and there
are enough tie stars when compared with deep ground-based imaging.
However, since the spectral response of the HST/STIS CCD is so broad,
extended objects with color gradients will have different apparent
relative locations when compared with our deep ground-based $R$-band
images.  As such, in choosing astrometric tie objects, we pay
particular attention to choosing objects which appear compact
(half-light radii $\ale$ 0\arcsec.3) on the STIS image.

\subsection{Field Distortion}
Optical field distortion is another source of potential error in
astrometric calibration.  Without correcting for distortion in STIS,
the maximum distortion displacement (on the field edges) is $\sim 35$
mas \citep{mb97}.  This distortion is corrected to a precision at the
sub-milliarcsec level on individual STIS exposures with IRAF/DITHER
\citep{mb97}. \citet{mb97} also found that the overall plate scale
appears to be quite stable with r.m.s.~changes at the 0.1\% level.  We
confirmed this result by comparing two epochs of imaging on GRB 990510
and GRB 970508 which span about 1 year. The relative plate scale of
the geometric mapping between final reductions was unity to within
0.03\%.

We do not correct for optical field distortion before mapping
ground-based images to HST.  While there may be considerable distortion
($\sim $ few $\times 100$ mas) across whole ground-based CCD images, these
distortions are correlated on small scales.  Therefore, when mapping a
50 $\times$ 50 arcsec$^2$ portion of a Keck image with an HST image,
the intrinsic differential distortions in the Keck image tend to be
small ($\ale 30$--50 mas).  Much of the distortion is accounted for in
the mapping by the higher-order terms of the fit, and any residual
differential distortions simply add scatter to the mapping
uncertainties.

\section{Derivation of the probability histogram (PH)}
\label{sec:osh-derive}

\def\xp{{\ensuremath{x}}}
\def\yp{{\ensuremath{y}}}
\def\rp{{\ensuremath{r}}}
\def\dxp{{\ensuremath{{\rm d}}\ensuremath{\xp}}}
\def\dyp{{\ensuremath{{\rm d}}\ensuremath{\yp}}}

Histogram binning is most informative when there are many more data
points than bins and the bin sizes are much larger than the errors on
the individual measurements.  Unfortunately, the set of GRB offsets is
contrary to both these requirements. We require a method to display
the data as in the traditional histogram, but where the errors on the
measurements are accounted for.  Instead of representing each
measurement as a $\delta$-function, we will represent each measurement
as a probability distribution as a function of offset.

What distribution function is suitable for offsets?  When the offset
is much larger than the error, then the probability that the burst
occurred at the measured displacement should approach a
$\delta$-function.  When the offset is much larger than zero, then the
probability distribution should appear essentially Gaussian (assuming
the error on the measurement is Gaussian).  However, when the observed
offset is small and the error on the measurement non-negligible with
respect to the observed offset, the probability distribution is
decidedly non-Gaussian since the offset is a positive quantity.  The
distribution we seek is similar to the well-known Rice distribution
\citepcf{wax54}, only more general.

We derive the probability histogram (PH) as follows.  For each GRB
offset, $i$, we construct an individual probability distribution
function $p_i(r)\,$d$r$ of the host-normalized offset ($r_i$) of the
GRB given the observed values for $X_{0,i}$, $Y_{0,i}$ and host
half-light radius $R_{i,{\rm half}}$ and the associated uncertainties.
To simplify the notation in what follows, we drop the index $i$ and
let all parameters with small capitalization represent dimensionless
numbers; for example, the value $x_{0} = X_{0}/R_{\rm half}$,
where $R_{\rm half}$ is the host half-light radius.  Without loss of
generality, we can subsume (by quadrature summation) the uncertainties
in the host center, the astrometric transformation, and the GRB center
into the error contribution in each coordinate. We assume that these
statistical coordinate errors are Gaussian distributed with
$\sigma_{\xp}$ and $\sigma_{\yp}$ with, for example,
$$
\sigma_{\xp} = \frac{X_0}{R_{\rm half}} \sqrt{ \frac{\sigma_{X_0}^2} {X_0^2} +
  \frac{\sigma_{R_{\rm half}}^2}{R_{\rm half}^2}}.
$$
Therefore, we can construct the probability $p(\xp,\yp)\,\dxp \, \dyp$
of the true offset at some distance $\xp$ and $\yp$ from the measured
offset location ($x_0, y_0$):
\begin{equation}
p(\xp,\yp)\,\dxp \, \dyp = 
	\frac{1}{2\pi\sigma_{\xp}\,\sigma_{\yp}}\,
	\exp\left[-\frac{1}{2}\left(\frac{\xp^2}{\sigma_{\xp}^2}
	+ \frac{\yp^2}{\sigma_{\yp}^2}\right)\right]\dxp \, \dyp,
\label{eq:pxy}
\end{equation}
assuming the errors in the $\xp$ and $\yp$ are uncorrelated. This is a
good approximation since, while the astrometric mappings generally
include cross-terms in $X$ and $Y$, these terms are usually small.  If
$\sigma_\xp = \sigma_\yp$, then eq.~\ref{eq:pxy} reduces to the
Rayleigh distribution in distance from the observed offset, rather
than the host center.

The probability distribution about the host center is found with an
appropriate substitution for $\xp$ and $\yp$ in eq.~\ref{eq:pxy}.  In
Figure \ref{fig:proboff} we illustrate the geometry of the
problem. The greyscale distribution shows $p(\xp,\yp)\,\dxp \, \dyp$
about the offset point $x_0$ and $y_0$.  Let $\phi = \tan^{-1}
(y_0/x_0)$ and transform the coordinates in eq.~\ref{eq:pxy} using
$\psi = \phi + \theta$, $\xp = r\,\cos \psi - x_0$, and $\yp = r\,\sin
\psi - y_0$.  The distribution we seek, the probability that the true
offset lies a distance $r$ from the host center, requires a
marginalization of $\int_\psi p_i(r,\psi)\,{\rm d}r\,{\rm d} \psi$
over $\psi$,
\begin{eqnarray}
p_i(r)\,{\rm d}r &=& \int_\psi p_i(r,\psi)\,{\rm d}r\,{\rm d} \psi
       \nonumber \\
       &=&  \frac{J\, {\rm d}r}{2\pi\sigma_{\xp}\,\sigma_{\yp}}\,
	\int_{0}^{2\pi}
	\exp\left[-\frac{1}{2}\left(\frac{\xp(r,\psi)^2}{\sigma_{\xp}^2}
	+ \frac{\yp(r,\psi)^2}{\sigma_{\yp}^2}\right)\right]
       \,{\rm d} \psi\label{eq:pr1},
\end{eqnarray} 
finding $J = r$ as the Jacobian of the coordinate transformation.  In
general, equation \ref{eq:pr1} must be integrated numerically using
the observed values $x_0$, $y_0$, $\sigma_{\xp}$, and $\sigma_{\yp}$.
The solution is analytic, however, if we assume that $\sigma_\xp
\rightarrow \sigma_\rp$ and $\sigma_\yp \rightarrow \sigma_\rp$, so
that,
\begin{eqnarray}
p_i(r)\,{\rm d}r &\approx& \frac{r}{\pi \sigma_{\rp}^2}\,
   \exp\left[-\frac{r^2 + r_0^2}{2\sigma_\rp^2}\right] \,
	\int_{\theta = 0}^{\pi}\exp \left[\frac{r\,r_0\,\cos
   \theta}{\sigma_\rp}\right] {\rm d} \, \theta {\rm d}r  \nonumber \\
		 &\approx& \frac{r}{ \sigma_{\rp}^2}\,
   \exp\left[-\frac{r^2 + r_0^2}{2\sigma_{\rp}^2}\right] \, 
           I_0\left(\frac{r\,r_{0}}{\sigma^2_{\rp}}\right) {\rm d} r,
	\label{eq:pr2}
\end{eqnarray}
where $I_0(x)$ is the modified Besel function of zeroth order and $r_0
= \sqrt{x_0^2 + y_0^2}$. 

The equation \ref{eq:pr2} is readily recognized as the Rice
distribution and is often used to model the noise characteristics of
visibility amplitudes in interferometry; visibility amplitudes, like
offsets, are positive-definite quantities.  Only when $\sigma_\xp =
\sigma_\yp = \sigma_r$ is the probability distribution exactly a Rice
distribution, which is usually the case for interferometric
measurements since the real and imaginary components of the fringe phasor
have the same r.m.s.

Equation \ref{eq:pr1} is a generalized form of the Rice distribution
but can be approximated as a Rice distribution by finding a suitable
value for $\sigma_r$.  We find that by letting,
\begin{equation}
\sigma_\rp = \frac{1}{r_0}\sqrt{ \left(x_0\, \sigma_{\xp}\right)^2 +
	\left(y_0\, \sigma_{\yp}\right)^2},
\label{eq:sigmar}
\end{equation}
equation \ref{eq:pr2} approximates (to better than 30\%) the exact
form of the probability distribution in eq.~\ref{eq:pr1} as long as
$\sigma_{\xp} \ale 2\,\sigma_{\yp}$ (or vise versa). In Figure
\ref{fig:probexam} we show two example offset probability
distributions in exact and approximate form.  Note that $r_0 -
\sigma_r \le r \le r_0 - \sigma_r$ is not necessarily the 68\% percent
confidence region of the true offset since the probability
distribution is not Gaussian. The exact form is used to construct the
data representations in Figures
\ref{fig:offset-cum}--\ref{fig:sfr-comp1}.

\section{Testing the Robustness of the KS test}
\label{sec:robust}

How robust are the estimates of probabilities found comparing the
observed distribution and the predicted progenitor offset
distributions?  Since there are different uncertainties on each offset
measurement, the KS test is not strictly the appropriate statistic to
determine the likelihood that the observed distribution could be drawn
from the same underlying (predicted) distribution.  One possibility is
to construct synthetic sets of observed data from the model using the
observed uncertainties.  However, a small uncertainty (say 0.2 arcsec
in radius observed to be paired with an equally small offset) which is
randomly assigned to a large offset from a Monte Carlo distribution
has a different probability distribution then if assigned to a small
offset (since the distribution in $r$ is only physical for positive
$r$).  Instead, we approach the problem from the other direction by
using the data themselves to assess the range in KS statistics given
our data. We construct $k=1000$ synthetic cumulative physical offset
distributions using the smoothed probability offset distributions
$p_i(r)\,$d$r$ for each GRB.  As before $r$ is the offset in units of
host half-light radius.  For each simulated offset distribution $k$ we
find a set \{$r_i$\}$_k$ such that,
$$
P[0,1] = \frac{\int_0^{r_i}  p_i(l)\,dl}{\int_0^{\infty}
p_i(l)\,dl},
$$
where $P[0,1]$ is a uniform random deviate over the closed interval
[0,1].  In addition, since some of the host assignments may be
spurious chance superpositions, we use the estimate of $P_{\rm
chance}$ (\S \ref{sec:angoffs}; Table \ref{tab:offnorm}) to
selectively remove individual offsets from a given Monte Carlo
realization of the offset dataset. GRBs with relatively secure host
assignments remain in more realizations than those without.  So, for
instance, the offset of GRB 980703 ($P_{\rm chance}$ = 0.00045) is
used in all realizations but the offset of GRB 970828 ($P_{\rm
chance}$ = 0.07037) is retained in only 93\% of the synthetic
datasets.

We evaluate the KS statistic as above for each synthetic set and
record the result. Figure \ref{fig:sfr-comp1} depicts the cumulative
probability distribution compared with the simple exponential disk
model.  The inset of the figure shows the distribution of KS
statistics for the set of synthetic cumulative distributions
constructed as prescribed above.  In both cases, as expected, the {\it
observed} KS probability falls near the median of the synthetic
distribution.  The distribution of KS statistics is not significantly
affected by retaining all GRB offsets equally (that is, assuming
$P_{\rm chance}$ = 0.0 for every GRB offset).  In table \ref{tab:nsks}
we present the result of the Monte Carlo modeling.  Using this
distribution of KS statistics we can now assess the robustness of our
comparison result: given the data and their uncertainties, the
probability that the observed GRB offset distribution is the same as
the model distribution of star formation (exponential disk) is $P_{\rm
KS} \ge 0.05$ in 99.6\% of our synthetic datasets.

\singlespace

\begin{figure*}
\centerline{\psfig{file=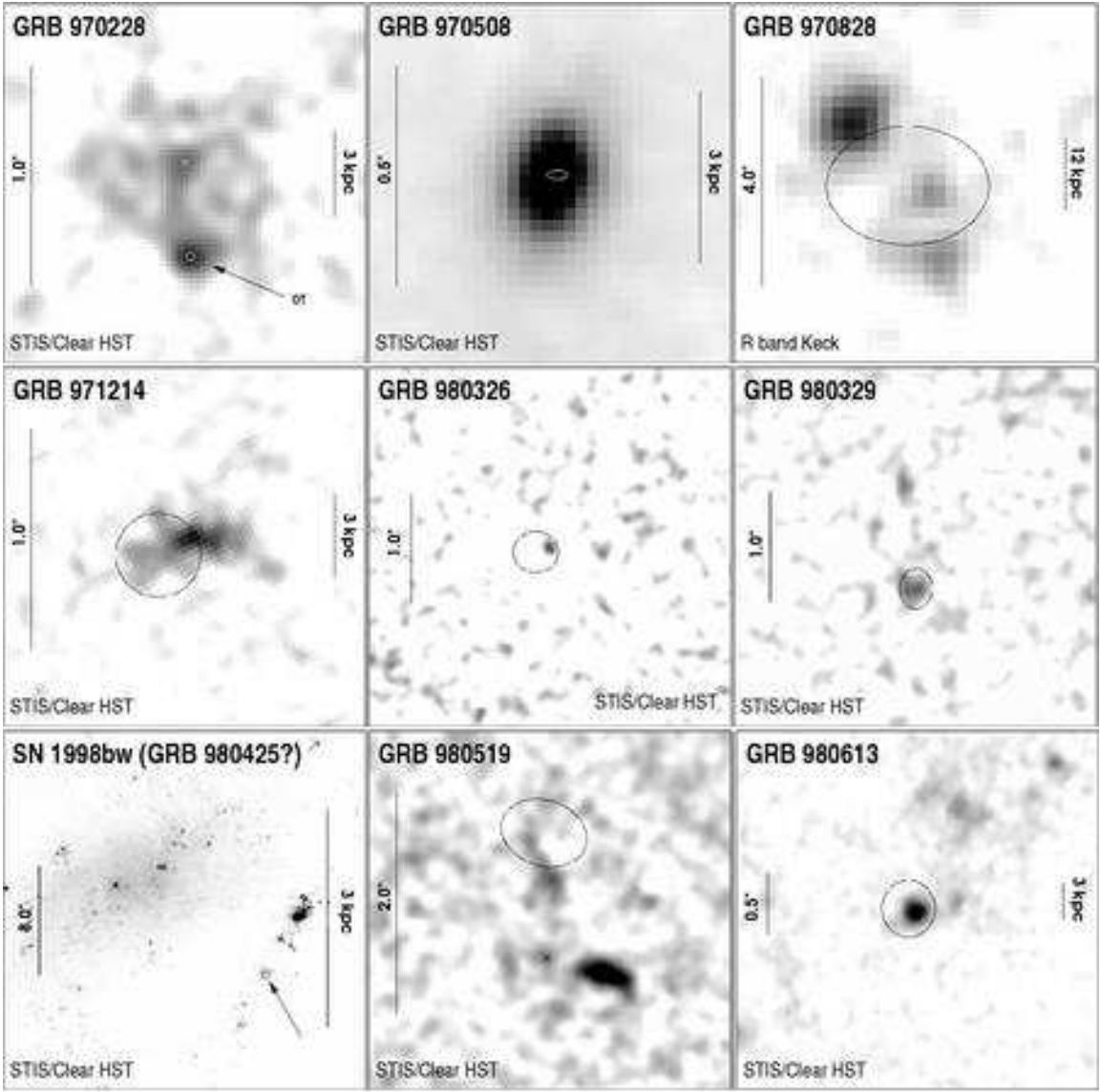,width=6in,angle=0}}
\caption[]{{\small The location of individual GRBs about their host
galaxies. The ellipse in each frame represents the 3-$\sigma$ error
contour for the location of the GRB as found in \S \ref{sec:indivoff}
and in Table \ref{tab:offsets}.  The angular scale of each image is
different and noted on the left hand side.  The scale and stretch was
chosen to best show both the detailed morphology of the host galaxy
and the spatial relationship of the GRB and the host.  The GRB
afterglow is still visible is some of the images (GRB 970228, GRB
991216).  In GRB 980425, the location of the associated supernova is
noted with an arrow. In all cases where a redshift is available for
the host or GRB afterglow we also provide a physical scale of the
region on the right hand side of each image.  For all images, North is
up and East is to the left. To
view the full-resolution image, please download the PDF (PS) version of the
paper at http://www.astro.caltech.edu/$\sim$jsb/Papers/offset.pdf (.ps.gz).}}
\label{fig:offset1}
\end{figure*}

\begin{figure*}
\figurenum{\ref{fig:offset1}}
\centerline{\psfig{file=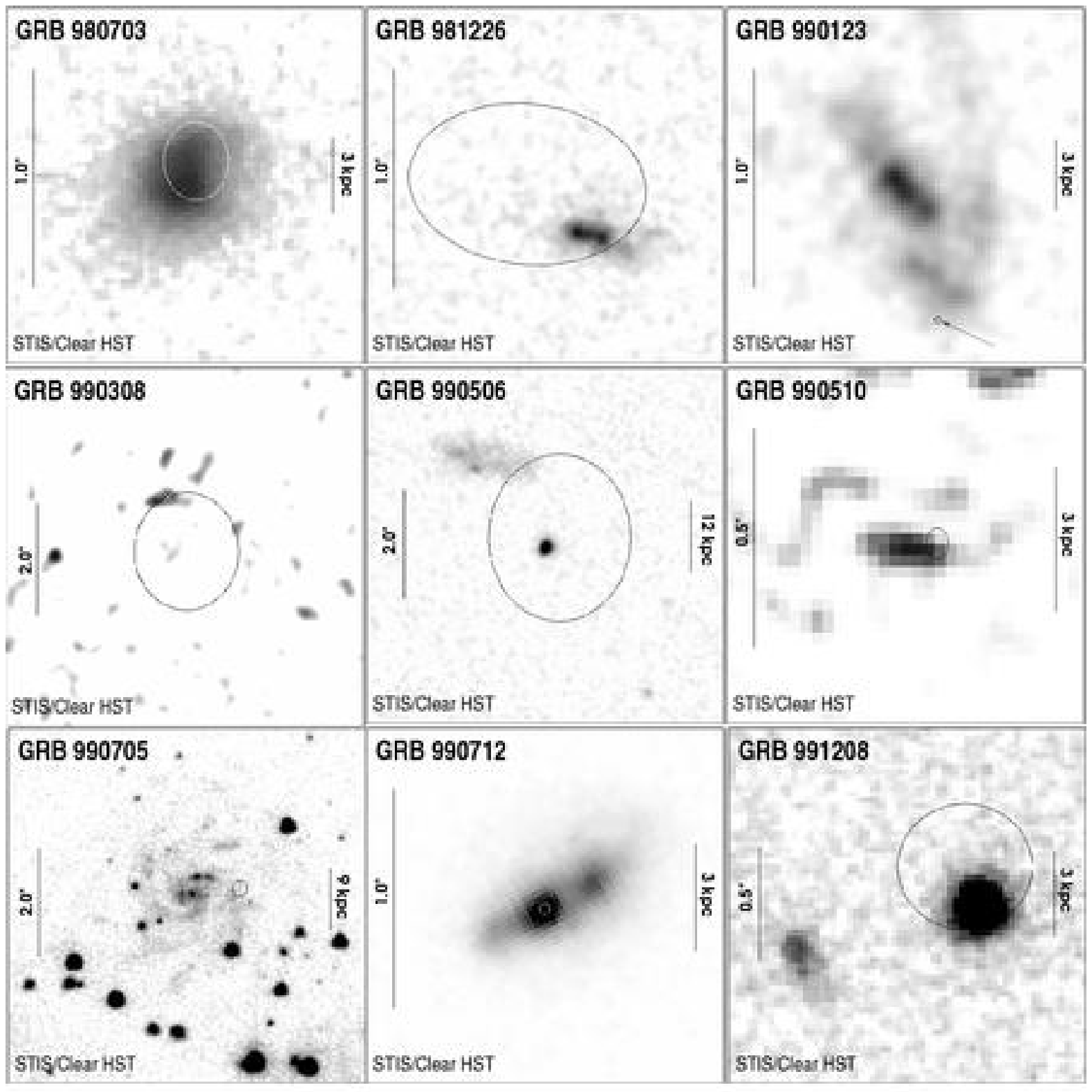,width=6in,angle=0}}
%
\caption[]{(cont.) The location of individual GRBs about their host
galaxies.}
\label{fig:offset2}
\end{figure*}

\begin{figure*}
\figurenum{\ref{fig:offset1}}
\centerline{\psfig{file=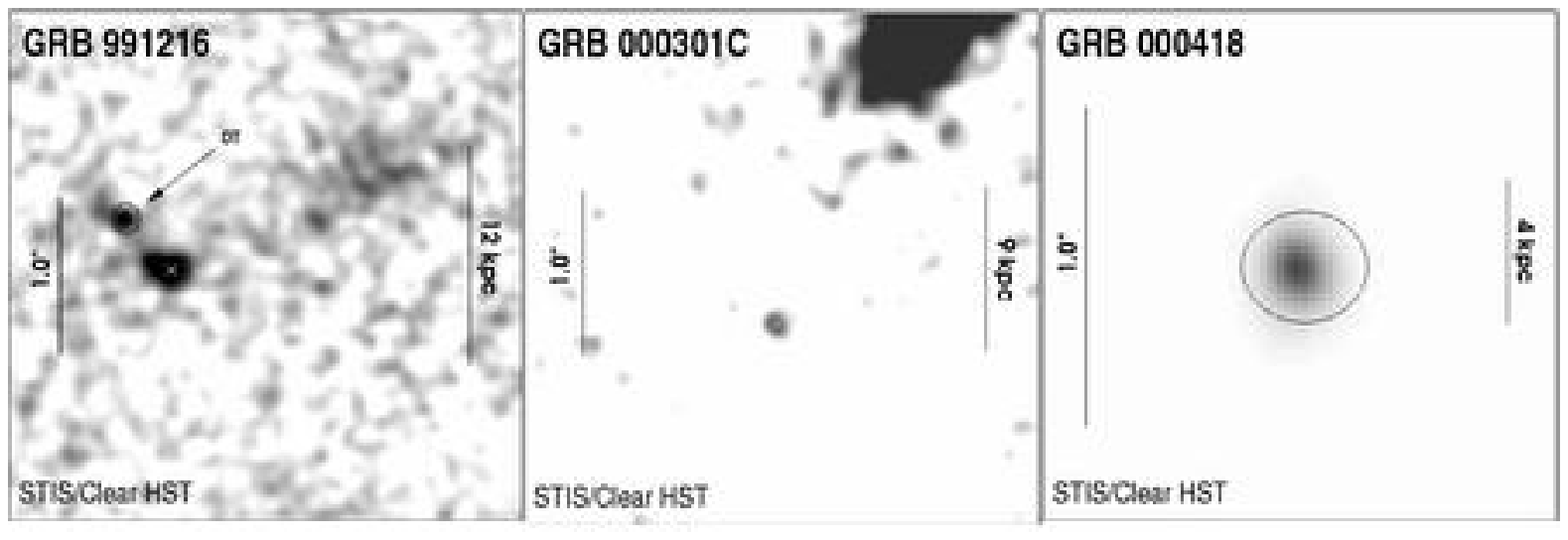,width=6in,angle=0}}
\caption[]{(cont.) The location of individual GRBs about their host
galaxies.}
\label{fig:offset3}
\end{figure*}

\begin{figure*}
\centerline{
\psfig{file=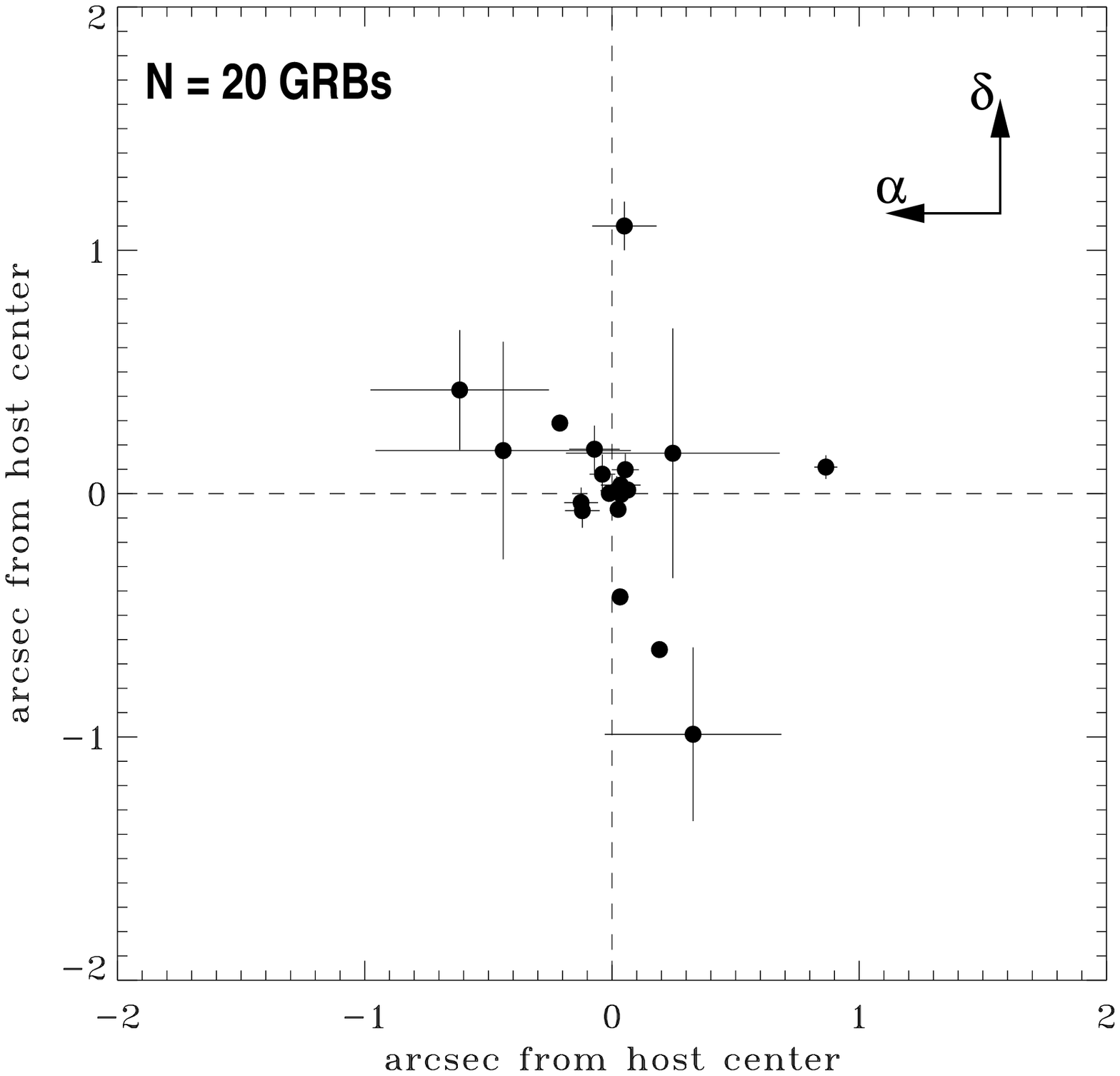,width=3.4in,angle=270}
\psfig{file=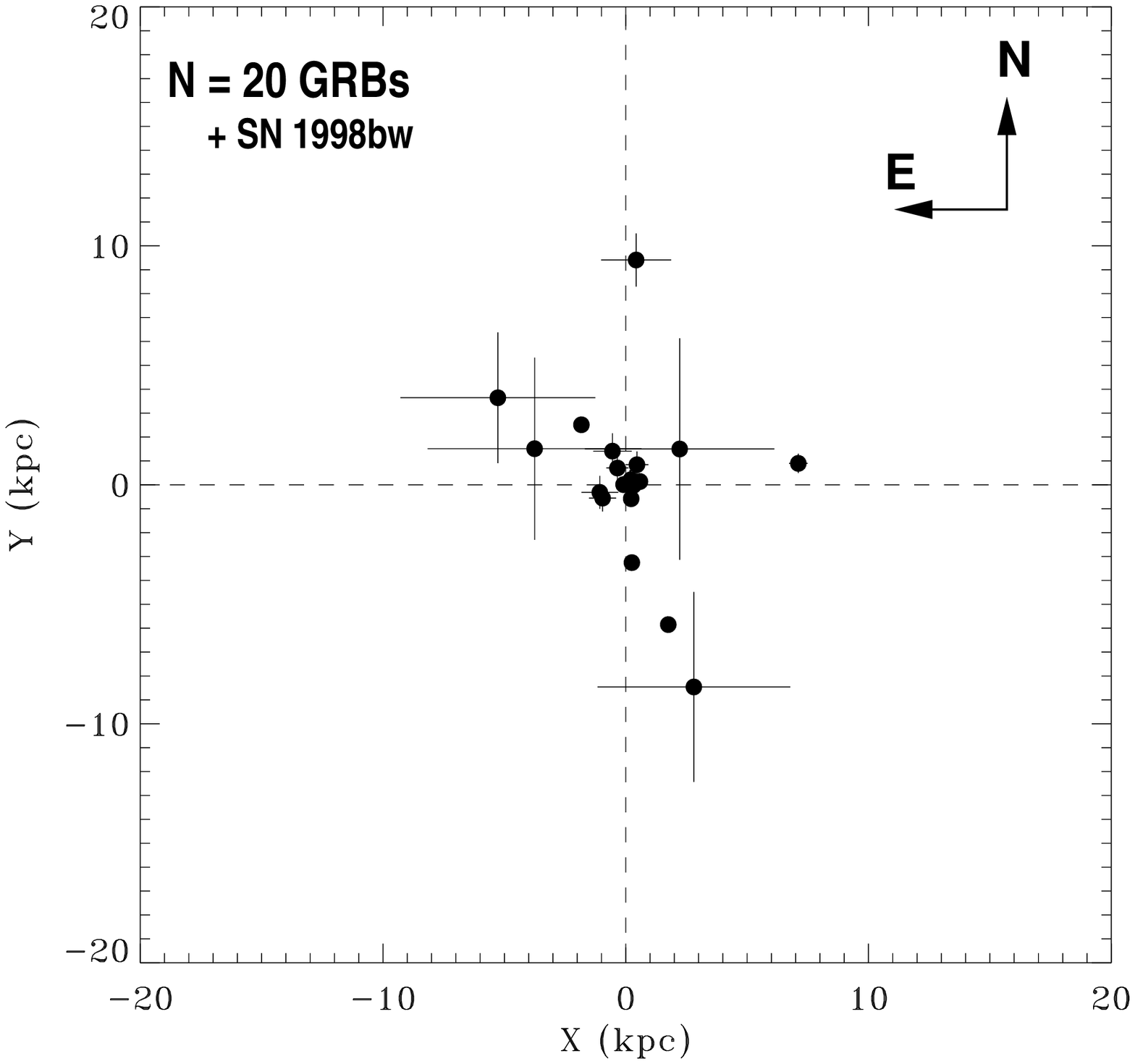,width=3.4in,angle=270}}
\caption[]{(left) The angular distribution of 20 gamma-ray bursts
about their presumed host galaxy. The error bars are 1 $\sigma$ and
reflect the total uncertainty in the relative location of the GRB and
the apparent host center.  The offset of GRB 980425 from its host is
suppressed for clarity since the redshift, relative to all the others,
GRB was so small. (right) The projected physical offset distribution
of 20 $\gamma$-ray bursts (now including SN1998bw/GRB 980425) about
their presumed host galaxies .  The physical offset is assigned
assuming $H_0 = 65$ km/s Mpc$^{-1}$, $\Lambda = 0.7$, and $\Omega_m =
0.3$ and assuming the GRB and the presumed host are at the same
redshift. Where no redshift has been directly measured a redshift is
assigned equal to the median redshift ($z = 0.966$) of all GRBs with
measured redshifts (see text).}
\label{fig:skyoff}
\end{figure*} 

\begin{figure*}[tbp]
\centerline{\psfig{file=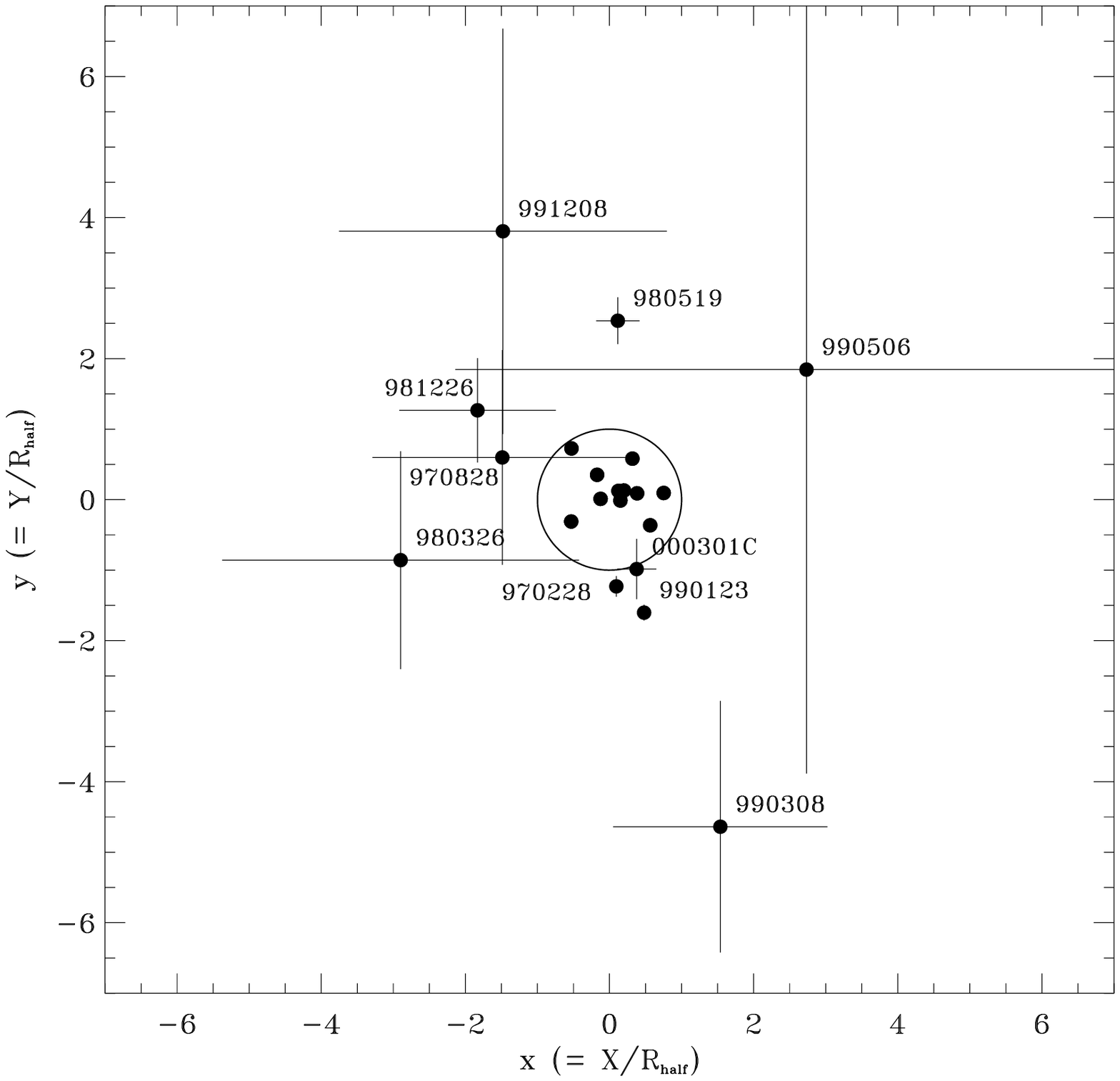,height=3.5in,angle=0}
\psfig{file=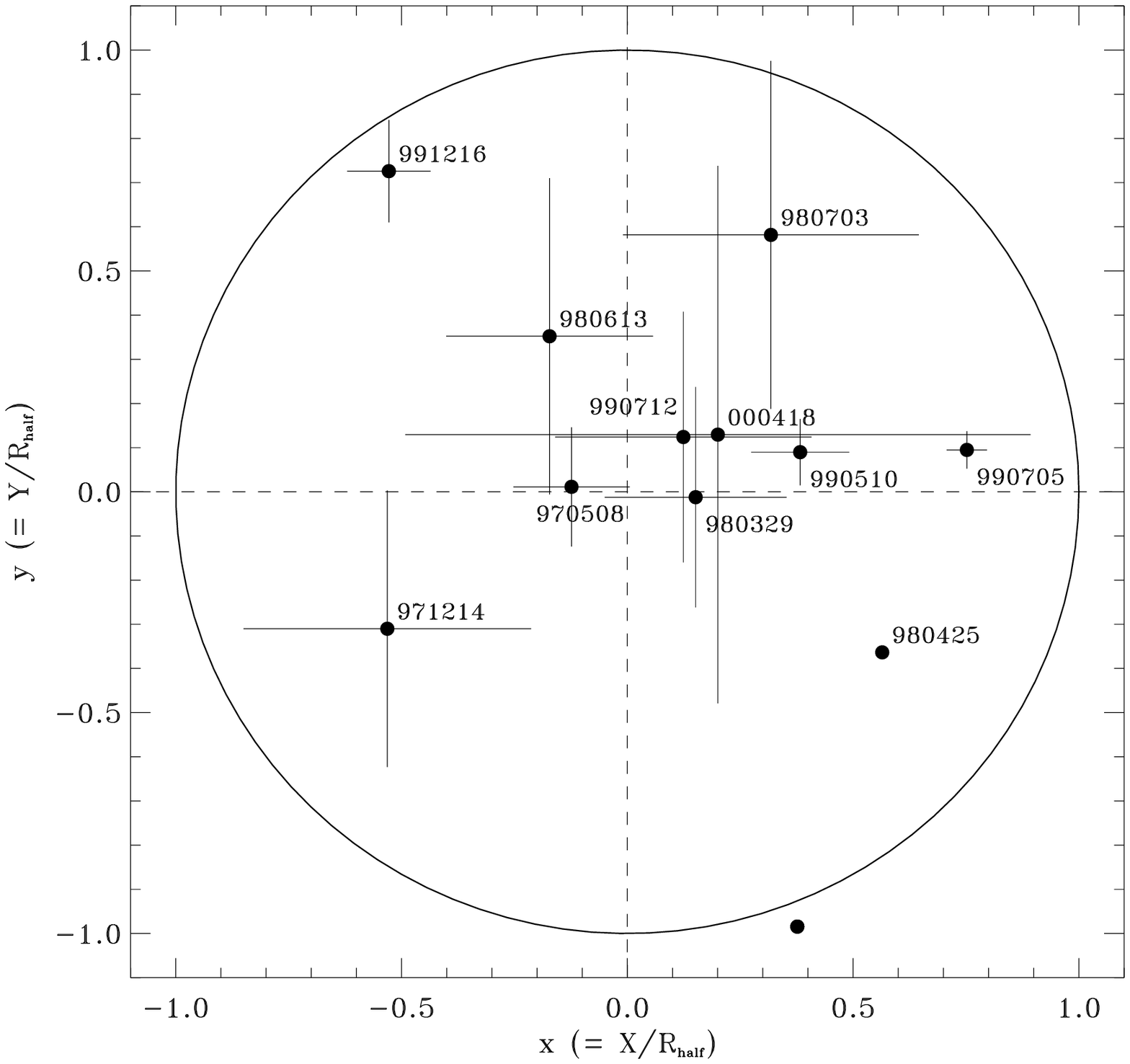,height=3.5in,angle=0}}
\caption[]{Host-normalized offset distribution.  The dimensionless
offsets are the observed offsets ($X_0, Y_0$) normalized by the host
half-light radius ($R_{\rm half}$) of the presumed host galaxy. See
text for an explanation of how the half-light radius is found.  The
1-$\sigma$ error bars reflect the uncertainties in the offset measure
and in the half-light radius.  As expected if GRBs occur where stars
are formed, there are 10 GRBs (plus 1998bw/GRB 980425) inside and 10
GRBs outside the half-light radius of their host.  (left) All GRBs
outside of one half-light radius (small circle) are labeled. (right)
All GRBs observed to be internal to one half-light radius are
labeled.}
\label{fig:hostnorm}
\end{figure*}

\begin{figure*}[tbp]
\centerline{\psfig{file=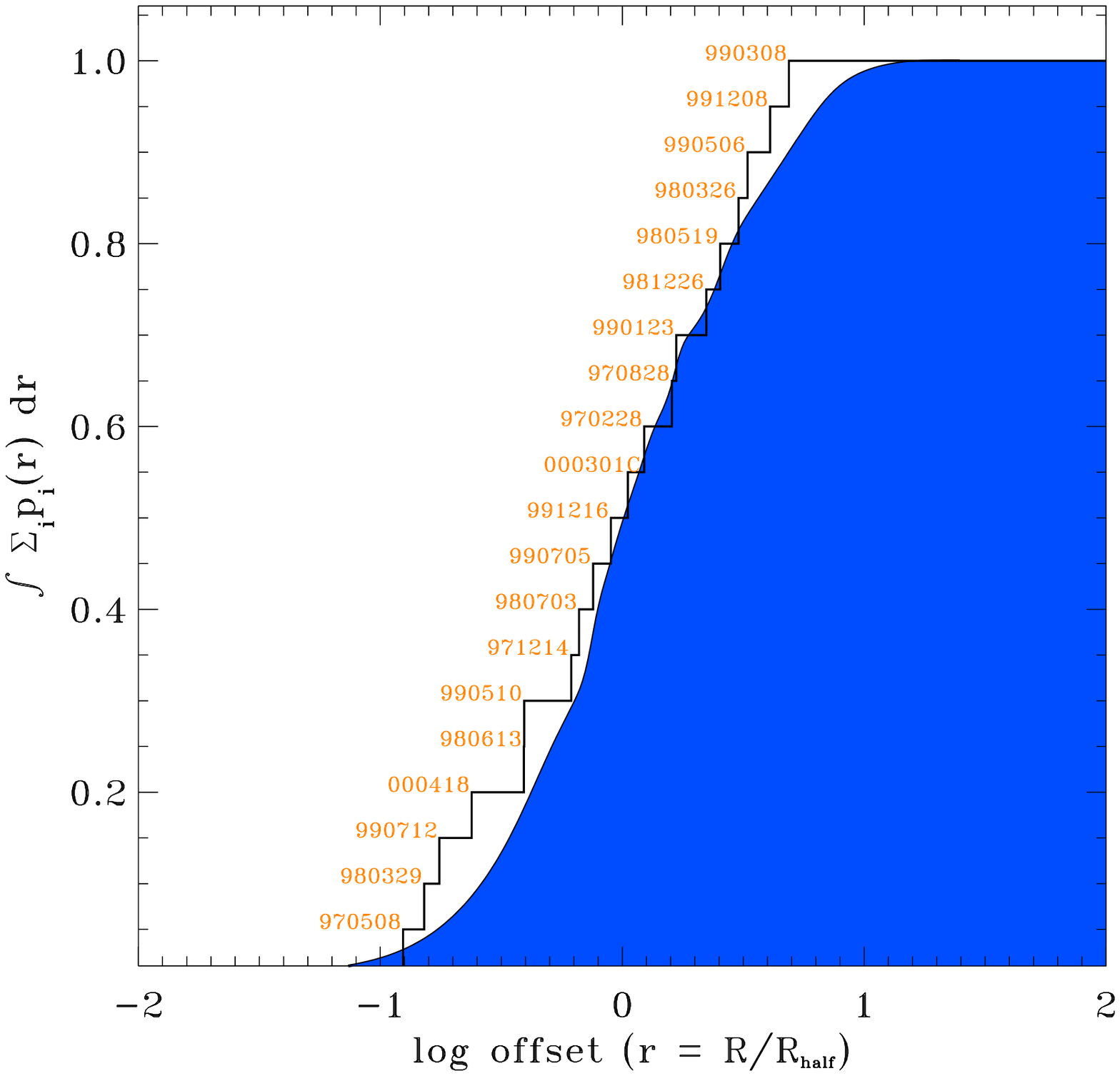,height=5.3in,angle=0}}
\caption[]{{\small The cumulative GRB offset distribution as a
function of host half-light radius.  The solid jagged line is the data
in histogram form. The smooth curve is the probability histogram (PH)
constructed with the formalism of Appendix \ref{sec:osh-derive} and is
the integral of the curve depicted in Figure \ref{fig:offset-log}. The
GRB identifications are noted alongside the solid histogram. In this
figure and in Figure \ref{fig:offset-log}, SN 1998bw/GRB 980425 has
not been included.}}
\label{fig:offset-cum}
\end{figure*}

\begin{figure*}[tbp]
\centerline{\psfig{file=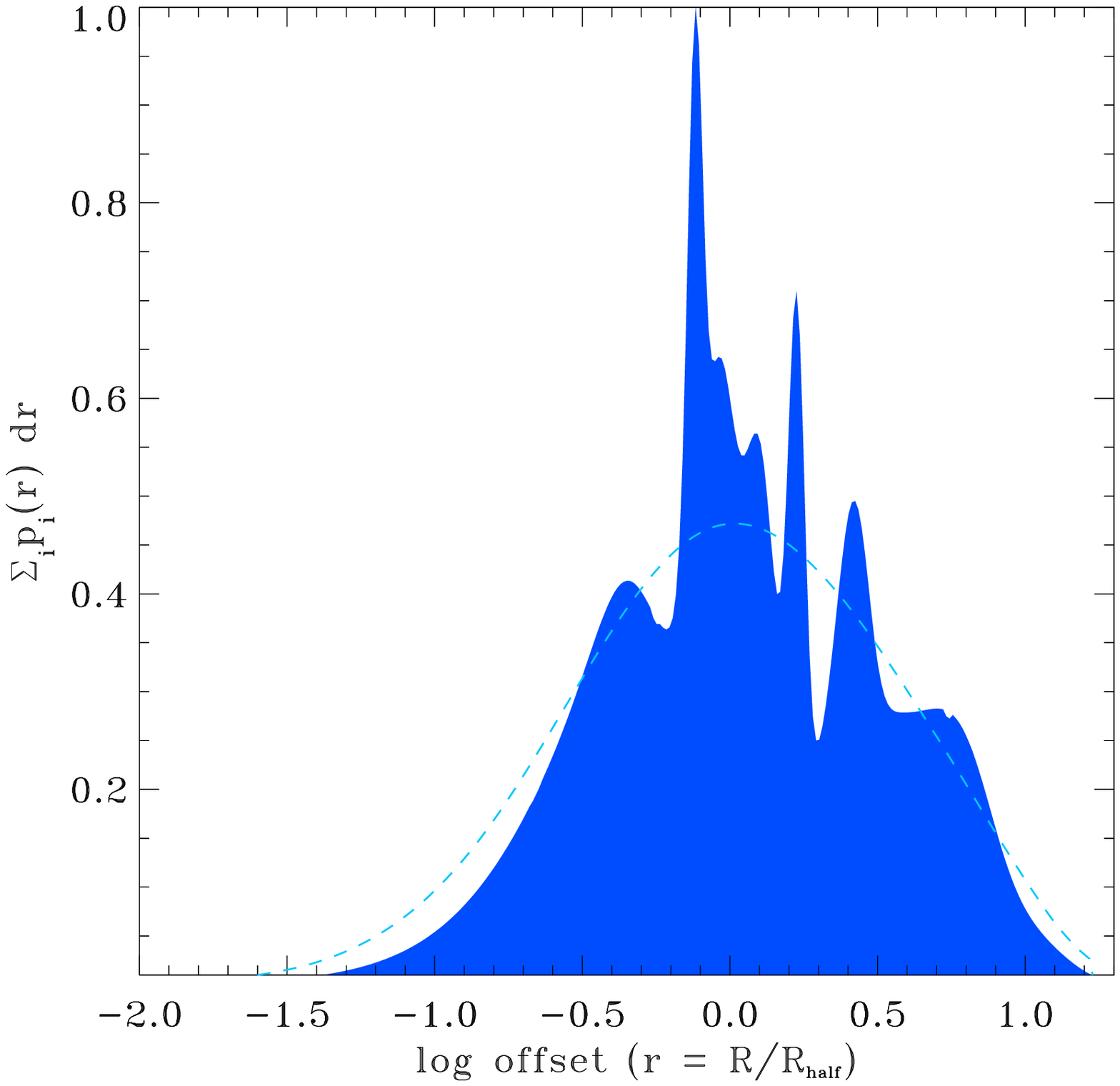,height=5.3in}}
\caption[]{{\small The GRB offset distribution as a function of
normalized galactocentric radius.  The normalized offset is $r =
R/R_{\rm half}$, where $R$ is the projected galactocentric offset of
the GRB from the host and $R_{\rm half}$ is the half-light radius of
the host. This distribution is essentially a smooth histogram of the
data but one which takes into account the uncertainties in the
measurements: the sharper peaks are due to individual offsets where
the significance ($r_0/\sigma_{r_0}$) of the offset is high. That is,
if a GRB offset is well-determined its contribution to the
distribution will appear as a $\delta$-function centered at $r =
r_0$.  There is no obvious host for this burst
and we choose the nearest galaxy detected as the host (see text).  The
dashed curve is the distribution under the blue (dark) curve but
smoothed with a Gaussian of FWHM = 0.7 dex in $r$.  Strikingly, the
peak of the probability is near one half-light radius, a qualitative
argument for the association of GRBs with massive star formation.  We
compare in detail this distribution with predicted progenitor
distributions in \S \ref{sec:compare}.}}
\label{fig:offset-log}
\end{figure*}

\begin{figure}
\centerline{\psfig{file=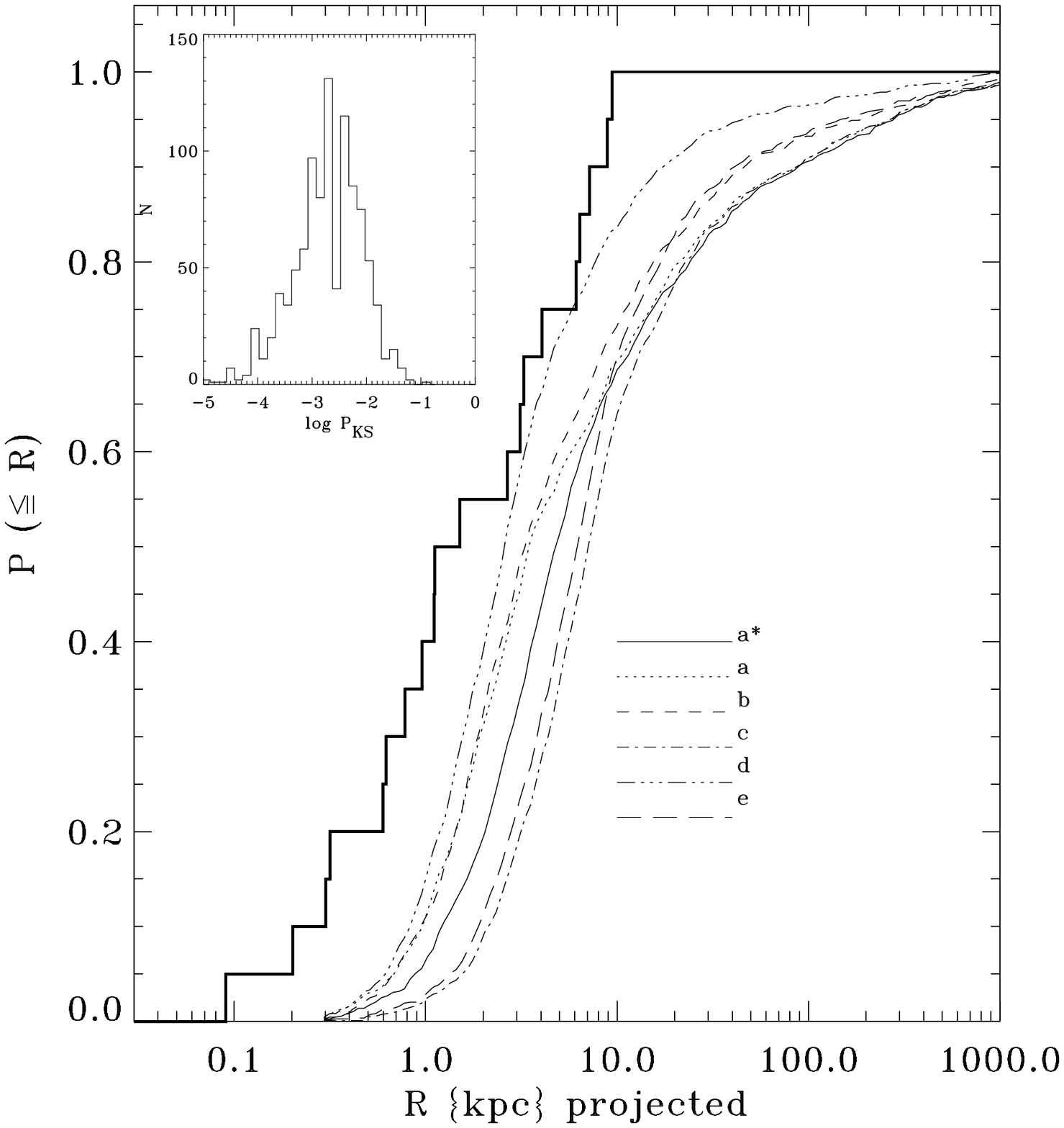,width=5.2in,angle=0}}
\caption[]{{\small Offset distribution of GRBs compared with delayed
merging remnant binaries (NS--NS and BH--NS) prediction. The models,
depicted as smooth curves, are the radial distributions in various
galactic systems that have been projected by a factor of 1.15 (see
text).  The letters denote the model distributions from Table 2 of
\citet{bsp99}; $a^*$ is the galactic model which we consider as the
most representative of GRB hosts galaxies ($v_{\rm circ} = 100$ km
s$^{-1}$, $r_{\rm break} = 1$ kpc, $r_e = 1.5$ kpc, $M_{\rm gal} = 9.2
\times 10^{9} M_\odot$).  The cumulative histogram is the observed
data set.  Inset is the distribution of KS statistics (based on the
maximum deviation from the predicted and observed distribution) of
1000 synthetic data sets compared with model $a^*$.  Even with
conservative assumptions (see text) the observed GRB distribution is
inconsistent with the prediction: in only 0.3\% of synthetic datasets
is $P_{\rm KS} \ge 0.05$. Instead, the collapsar/promptly bursting
remnant progenitor model appears to be a better representation of the
data (see Figure~\ref{fig:sfr-comp1}).}}
\label{fig:rem-comp1}
\end{figure}

\begin{figure}
\centerline{\psfig{file=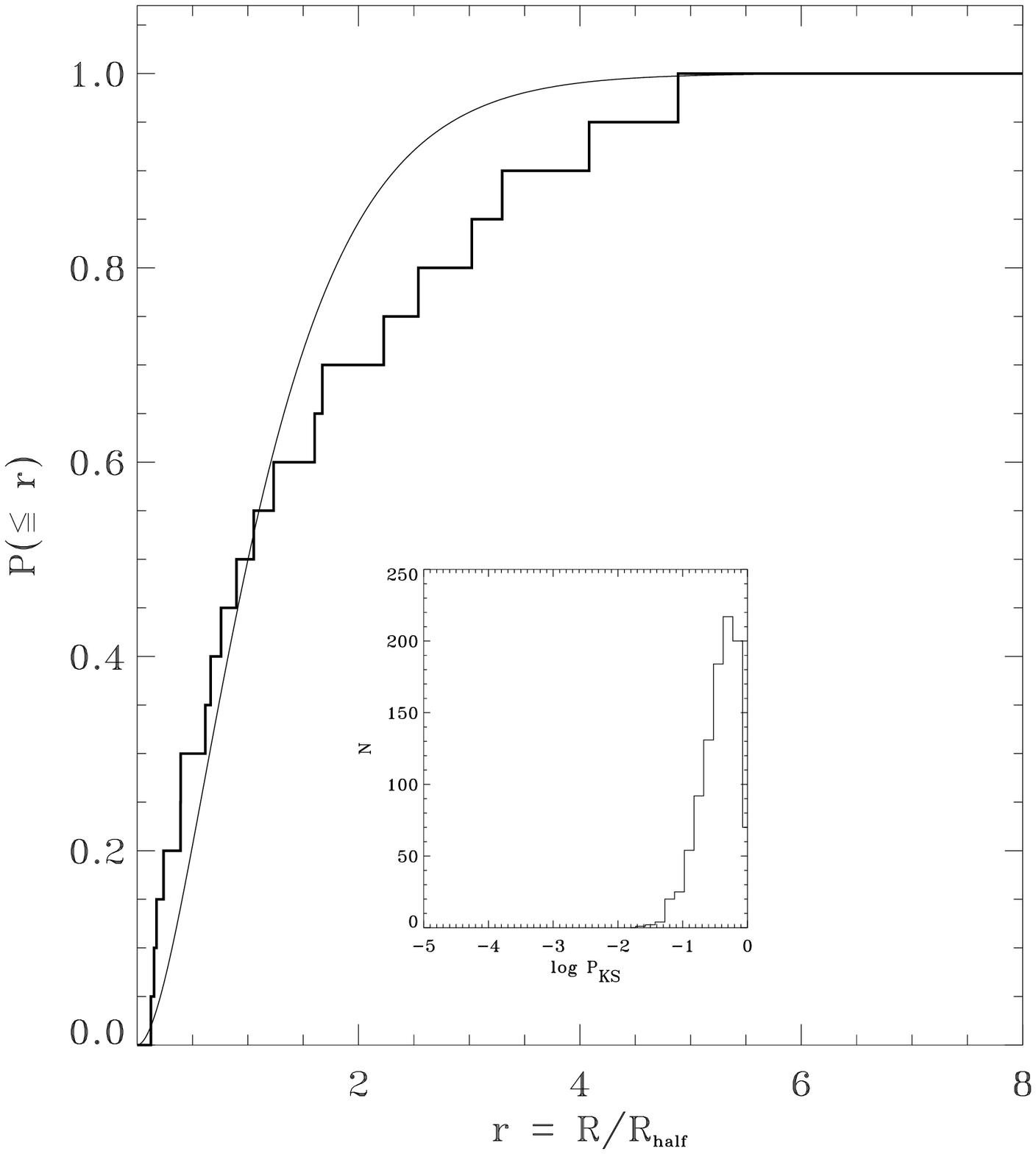,width=5.2in,angle=0}}
\caption[]{{\small Offset distribution of GRBs compared with host
galaxy star formation model. The model, an exponential disk, is shown
as the smooth curve and was chosen as an approximation to the
distribution of the location of collapsars and promptly bursting
remnant binaries (BH--He). The cumulative histogram is the observed
data set. Inset is the distribution of KS statistics (based on the
maximum deviation from the predicted and observed distribution) of
1000 synthetic data sets.  Since the observed KS statistic is near the
median in both cases, we are assured that errors on the measurements
do not bias the results of the KS test, and therefore the KS test is
robust. The observed GRB distribution provides a good fit to the model
considering we make few assumptions to perform the comparison.  In
reality the location of star formation in GRB hosts will be more
complex than a simple exponential disk model.}}
\label{fig:sfr-comp1}
\end{figure}

\begin{figure}
\centerline{\psfig{file=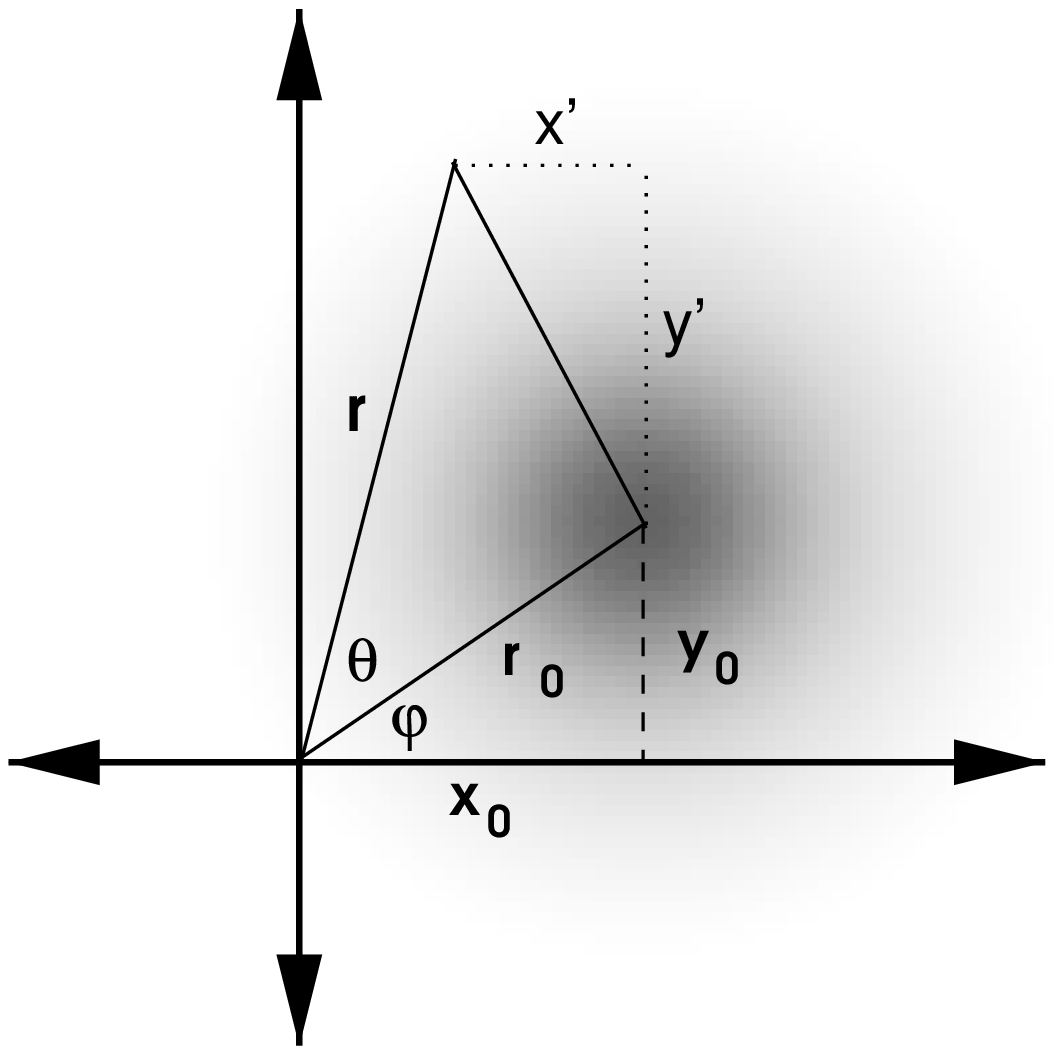,width=3.2in,angle=270}}
\caption[]{Geometry for the offset distribution probability calculation
in Appendix \ref{sec:osh-derive}. }
\label{fig:proboff}
\end{figure}
\begin{figure}
\centerline{\psfig{file=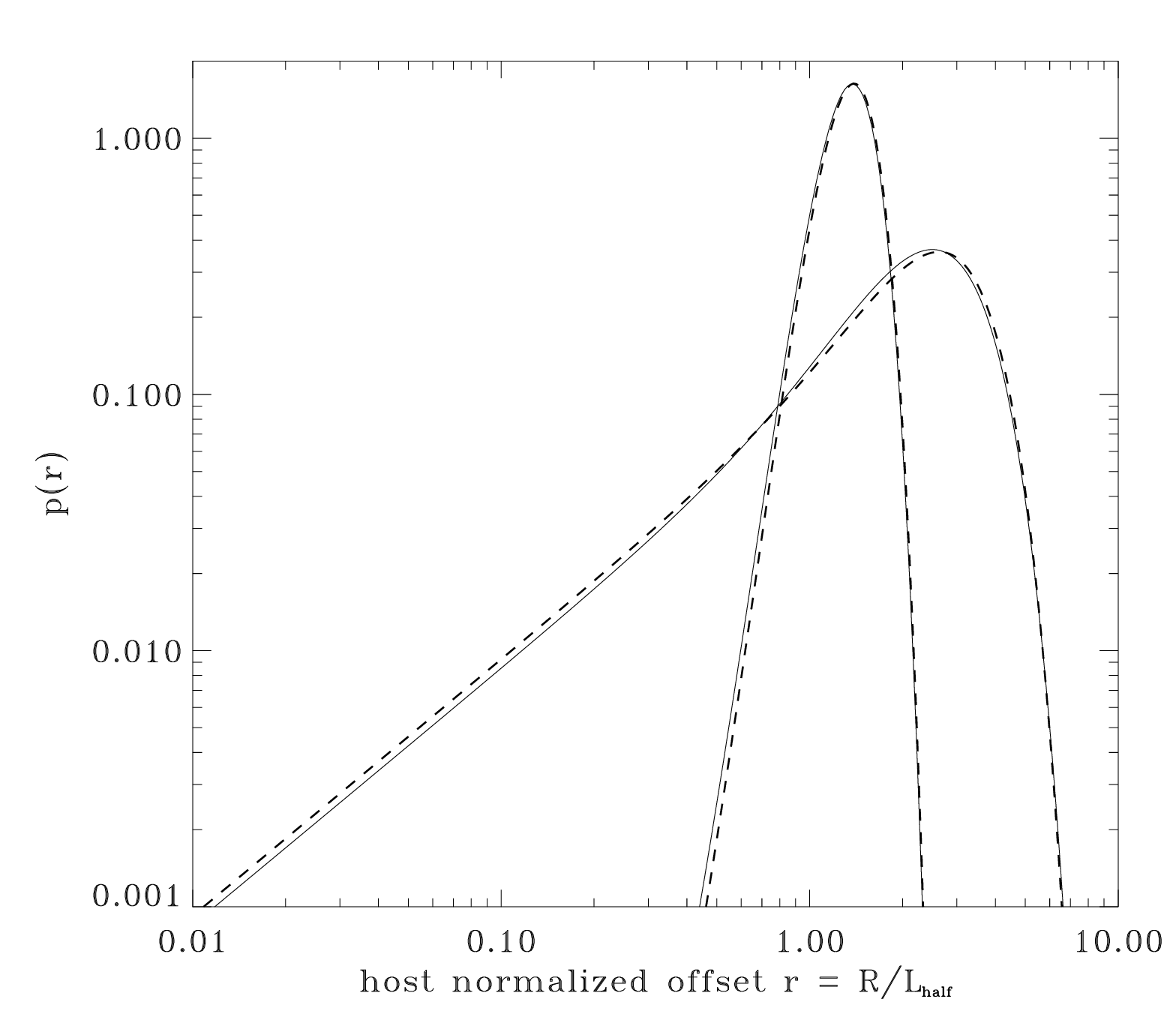,width=3.2in}}
\caption[]{Example offset distribution functions $p(r)$. Depicted are
two probability distribution curves for ($X_0$, $Y_0$, $\sigma_{X_0}$,
$\sigma_{Y_0}$, $R_{\rm half}$, $\sigma_{R_{\rm half}}$) =
[0\arcsec.033, 0\arcsec.424, 0\arcsec.034, 0.\arcsec034, 0\arcsec.31,
0\arcsec.05] (GRB 970228) and [0\arcsec.616, 0\arcsec.426,
0\arcsec.361, 0\arcsec.246, 0\arcsec.314, 0\arcsec.094] (GRB 981226)
for the lower and upper peaked distributions, respectively. The solid
line is the exact solution (eq.~\ref{eq:pr1}) and the dashed line is
the approximate solution (eq.~\ref{eq:pr2}). Here, as in the text, the
host-normalized offset $r = R/R_{\rm half}$, where $R$ is the
galactocentric offset of the GRB from the host and $R_{\rm half}$ is
the half-light radius of the host.}
\label{fig:probexam}
\end{figure}

\newpage

\def\me{a}
\def\vgg{1}
\def\fpt{2}
\def\fkn{3}
\def\pfb{4}
\def\hp{5}
\def\ggv{6}
\def\db{7}
\def\kdr{8}
\def\odk{9}
\def\ggvc{10}
\def\bkde{11}
\def\fvn{12}
\def\lgk{13}
\def\hth{14}
\def\gvv{15}
\def\hft{16}
\def\jha{17}
\def\dgk{18}
\def\bkdg{19}
\def\hfta{20}
\def\hap{21}
\def\hthf{22}
\def\bfk{23}
\def\bk{24}
\def\fkb{25}
\def\hta{26}
\def\bod{27}
\def\ssh{28}
\def\hftb{29}
\def\tbfk{30}
\def\htab{31}
\def\vgr{32}
\def\ffp{33}
\def\bkul{34}
\def\mpp{35}
\def\hah{36}
\def\sahu{37}
\def\fsg{38}
\def\fvh{39}
\def\fra{40}
\def\bdk{41}
\def\dcb{42}
\def\fvsc{43}
\def\umh{44}
\def\vff{45}
\def\fjh{46}
\def\fsg{47}
\def\fv{48}
\def\mhw{49}
\def\bdg{50}
\def\mfm{51}

\newpage
\singlespace

\begin{deluxetable}{llrrrrrrll}
\rotate
\tabletypesize{\scriptsize}
\tablewidth{8.4in}
\tablecaption{GRB Host and Astrometry Observing Log\label{tab:offset-log}}
\tablehead{
\colhead{Name} &
\colhead{Teles./Instr./Filter} &
\colhead{Date} &
\colhead{$\alpha$ (J2000)} &
\colhead{$\delta$ (J2000)} &
\colhead{Exp.} &
\colhead{$\Delta t$} &
\colhead{Level} &
\colhead{Refs.} \\
\colhead{(1)} &
\colhead{(2)} &
\colhead{(3)} &
\multicolumn{2}{c}{(4)} &
\colhead{(5)} &
\colhead{(6)} &
\colhead{(7)} &
\colhead{(8)}}
\startdata

GRB 970228\ldots &HST/STIS/{\tt O49001040} & 4.75 Sep 1997 & 05 01 46.7 & +11 46 54  & 4600  & 189 & self-HST              & \vgg, \fpt \\

GRB 970508\ldots &HST/STIS/{\tt O41C01DIM} & 2.64 Jun 1997 & 06 53 49.5 & +79 16 20  & 5000  & 25  & HST$\rightarrow$HST   & \fkn, \pfb \\*
                 &HST/STIS/{\tt O4XB01I9Q} & 6.01 Aug 1997 &             &           & 11568 & 89  &                       & \hp \\

GRB 970828\ldots & Keck/LRIS/$R$-band    & 19.4 Jul 1998     & 18 08 34.2 & +59 18 52  & 600   & 325 & RADIO$\rightarrow$OPT  & \ggv, \db \\

GRB 971214\ldots & Keck/LRIS/$I$-band    & 16.52  Dec 1997   & 11 56 26.0 & +65 12 00  & 1080  & 1.5 & GB$\rightarrow$HST  & \kdr \\*
                 & HST/STIS/{\tt O4T301040}& 13.27 Apr 1998&            &            & 11862 & 119 &                       & \odk \\

GRB 980326\ldots & Keck/LRIS/$R$-band    & 28.25 Mar 1998    & 08 36 34.3 & $-$18 51 24& 240   & 1.4 & GB$\rightarrow$GB &\ggvc,\bkde \\*
                 & Keck/LRIS/$R$-band    & 18.50 Dec 1998    &            &            & 2400  & 267 &                       & \bkde \\
                 & HST/STIS/{\tt O4T301040}& 31.80 Dec 2000 &            &            & 7080 & 1010 &                       & \fvn \\

GRB 980329\ldots & Keck/NIRC/$K$-band    & 2.31 Apr 1998    & 07 02  38.0 &  +38 50 44 & 2520  & 4.15 & GB$\rightarrow$GB  & \lgk,  \me \\
                 & Keck/ESI/$R$-band     & 1.41 Jan 2001    &  &  & 6600  & 1009 &   & \me \\
                 & HST/STIS/{\tt O65K22YXQ}& 27.03 Aug 2000 &            &            & 8012 & 884 &                       & \hth \\

SN 1998bw  \ldots &NTT/EMMI/$I$-band       & 4.41 May 1998   & 19 35 03.3  &$-$52 50 45& 120   & 8.5 & GB$\rightarrow$HST   & \gvv \\*
\phantom{~~~}(GRB 980425?)     &HST/STIS/{\tt O65K30B1Q} & 11.98 Jun 2000  &          &           & 1185  & 778 &                       & \hft \\

GRB 980519\ldots & P200/COSMIC/$R$-band   & 20.48 May 1998   & 23 22 21.5  &  +77 15 43& 480   & 1.0 &GB$\rha$GB$\rha$HST & \jha, \dgk \\*
                 & Keck/LRIS/$R$-band     & 24.50 Aug 1998   &             &           & 2100  & 97  &                    & \bkdg \\*
                 & HST/STIS/{\tt O65K41IEQ} & 7.24 Jun 2000&             &           & 8924  & 750 &                       & \hfta \\

GRB 980613\ldots  & Keck/LRIS/$R$-band     & 16.29 Jun 1998  & 10 17 57.6  & +71 27 26 & 600   & 3.1 &GB$\rightarrow$GB$\rha$HST  & \hap, \me  \\*
                  & Keck/LRIS/$R$-band     & 29.62 Nov 1998  &             &           & 900   & 169 &                       & \me  \\
& HST/STIS/{\tt O65K51ZZQ} & 20.31 Aug 2000 &             &           & 5851  & 799 &                       & \hthf \\

GRB 980703\ldots  & Keck/LRIS/$R$-band     & 6.61 Jul 1998   & 23 59 06.7  & +08 35 07 & 600   & 3.4 &GB$\rightarrow$HST   & \bfk \\*
            & HST/STIS/{\tt O65K61XTQ} & 18.81 Jun 2000  &               &           & 5118  & 717 &                       & \bk \\

GRB 981226\ldots  & Keck/LRIS/$R$-band     & 21.57 Jun 1999  & 23 29 37.2 & $-$23 55 54 & 3360 & 177 &RADIO$\rha$OPT(GB)$\rha$HST  & \fkb  \\*
            & HST/STIS/{\tt O65K71AXQ} & 3.56  Jul 2000  &              &             & 8265 & 555 &                      & \hta  \\

GRB 990123\ldots  & HST/STIS/{\tt O59601060} & 9.12 Feb 1999 & 15 25 30.3 & +44 45 59 & 7200 & 16.7 & self-HST & \bod \\
            
GRB 990308\ldots  & Keck/LRIS/$R$-band       & 19.26 Jun 1999  & 12 23 11.4 & +06 44 05& 1000& 103 &  GB$\rha$GB$\rha$HST & \ssh  \\*
                  & HST/STIS/{\tt O65K91E6Q} & 19.67 Jun 2000  &            &          & 7782& 470 &                      & \hftb \\

GRB 990506\ldots  & Keck/LRIS/$R$-band       & 11.25 Jun 1999  & 11 54 50.1 & $-$26 40 35& 1560& 36 & RADIO$\rha$OPT(GB)$\rha$HST  & \tbfk \\*
                  & HST/STIS/{\tt O65KA1UYQ} &24.55 Jun 2000 &            &            & 7856& 415  &                       & \htab \\

GRB 990510\ldots  & HST/STIS/{\tt O59273LCQ} & 17.95 Jun 1999 & 13 38 07.7& $-$80 29 49& 7440& 39   & HST$\rightarrow$HST  & \vgr, \ffp \\* 
                  & HST/STIS/{\tt O59276C7Q} & 29.45 Apr 2000 &           &            & 5840& 355  &                      & \bkul \\

GRB 990705\ldots  & NTT/SOFI/$H$-band          & 5.90  Jul  1999 & 05 09 54.5& $-$72 07 53& 1200& 0.23 & GB$\rha$GB$\rha$HST   & \mpp \\*
                  & VLT/FORS1/$V$-band     & 10.40 Jul  1999 &           &            & 1800& 4.7  &                            & \mpp \\*
                  & HST/STIS/{\tt O65KB1G2Q} & 26.06 Jul  2000 &           &            & 8792& 386  &                       & \hah \\

GRB 990712\ldots  & HST/STIS/{\tt O59262VEQ} & 29.50 Aug 1999  & 22 31 53.1& $-$73 24 28& 8160 & 48 & HST$\rha$HST & \sahu, \fsg \\*
                   & HST/STIS/{\tt O59274BNQ} & 24.21 Apr 2000  &           &            & 3720 & 287 &            & \fvh \\

GRB 991208\ldots  & Keck/NIRSPEC/$K$-band  & 16.68 Dec 1999 & 16 33 53.5  & +46 27 21 & 1560   & 8.5 & GB$\rha$GB$\rha$HST & \fra, \bdk \\*
                  & Keck/ESI/$R$-band      & 4.54 Apr 2000  &             &             & 1260 & 118 &                & \me \\* 
         & HST/STIS/{\tt O59266ODQ} & 3.58 Aug 2000  &           &            & 5120 & 239 &            & \fvsc \\

GRB 991216\ldots  & Keck/ESI/$R$-band        & 29.41 Dec 1999  & 05 09 31.2  & +11 17 07   & 600 &13&  GB$\rha$GB$\rha$HST  & \umh, \me \\*
	          & Keck/ESI/$R$-band         & 4.23 Apr 2000  &             &             & 2600 & 110 &                    & \me \\*
	          & HST/STIS/{\tt O59272GIQ} & 17.71 Apr 2000&             &             & 9440 & 123 &                    & \vff \\

GRB 000301C\ldots & HST/STIS/{\tt O59277P9Q} & 6.22 Mar 2000 & 16 20 18.6 & +29 26 36 & 1440 & 4.8 & HST$\rha$HST & \fjh, \fsg \\
                  & HST/STIS/{\tt O59265XYQ} & 25.86 Feb 2001 &            &
        & 7361 &  361  &  & \fv \\

GRB 000418\ldots  & Keck/ESI/$R$-band & 28.41 Apr 2000 & 12 25 19.3 & +20 06 11   &  300 & 10 & GB$\rha$HST  & \mhw, \bdg \\*
                  & HST/STIS/{\tt O59264Y6Q}           & 4.23  Jun 2000  &            &    & 2500 & 47 &                 & \mfm  

\enddata \tablecomments{(2) Telescopes: HST = {\it Hubble Space
Telescope} Keck = W.~M.~Keck 10 m Telescope II, Mauna Kea, Hawaii,
P200 = Hale 200-inch Telescope at Palomar Observatory, Palomar
Mountain, California, NTT = European Space Agency 3.5 m New Technology
Telescope, Chile, VLT = Very Large Telescope UT-1 (``Antu'');
Instruments: STIS \citep{kwb+98}, ESI \citep{em98}, LRIS
\citep{occ+95}, COSMIC \citep{kds+98}, NIRSPEC \citep{mbb+98}, SOFI
\citep{fbm+98}, FORS1 \citep{nsb+97}; Filter: all ground-based
observations are listed in standard bandpass filters while the
HST/STIS images (used for astrometry) are all in Clear Mode.  The last
dataset of the HST visit is listed. (3) Observation dates in Universal
Time (UT) corresponding to the start of the last observation in the
dataset. (4) Position ($\alpha$: hours, minutes, seconds and $\delta$:
degrees, arcminutes, and arcseconds) of the GRB. (5) Total exposure
time in seconds. (6) Time in days since the trigger time of the
GRB. (7) The comment denotes the astrometric level as in \S
\ref{sec:astlevels}. (8) Reference to the first presentation of the
given dataset. If two references appear on a given line then the first
is a reference to the position of the GRB.}

\tablerefs{\me.~This paper; \vgg.~\citet{vgg+97};
\fpt.~\citet{fpt+99}; \fkn.~\citet{fkn+97}; \pfb.~\citet{pfb+98a};
\hp.~\citet{fp98}; \ggv.~\citet{ggv+98d}; \db.~\citet{d00};
\kdr.~\citet{kdr+98}; \odk.~\citet{odk+98}; \ggvc.~\citet{ggv+98c};
\bkde.~\citet{bkd+99}; \fvn.~\citet{fvn01}; \lgk.~\citet{lgk+98};
\hth.~\citet{hth+00}; \gvv.~\citet{gvv+98}; \hft.~\citet{hft+00};
\dgk.~\citet{dgk+98}; \bkdg.~\citet{bkdg+98}; \hfta.~\citet{hfta+00};
\hap.~\citet{hap+98}; \hthf.~\citet{hth+00}; \bfk.~\citet{bfk+98};
\bk.~\citet{bk+00}; \fkb.~\citet{fkb+99}; \hta.~\citet{hta+00};
\bod.~\citet{bod+99}; \ssh.~\citet{ssh+99}; \hftb.~\citet{hft+00b};
\tbfk.~\citet{tbf+00}; \htab.~\citet{htab+00}; \vgr.~\citet{vgr+99};
\ffp.~\citet{ffp+99}; \bkul.~\citet{blo00}; \mpp.~\citet{mpp+00};
\hah.~\citet{hah+00}; \sahu.~\citet{sah+00}; \fsg.~\citet{fsg+00};
\fvh.~\citet{fvh+00}; \fra.~\citet{frail+00}; \bdk.~\citet{bdk+00};
\dcb.~\citet{dcb+00}; \fvsc.~\citet{fvsc00}; \umh.~\citet{umh+00};
\vff.~\citet{vff+00}; \fjh.~\citet{fjh+00}; \fsg.~\citet{fsg+00};
\fv.~\citet{fv+01}; \mhw.~\citet{mhw+00}; \bdg.~\citet{bdg+00};
\mfm.~\citet{mfm+00}}

\end{deluxetable}

\newpage

\begin{deluxetable}{lccccccccc}
\tabletypesize{\small}
\rotate
\tablewidth{9in}
\tablecaption{Measured angular offsets and physical projections\label{tab:offsets}}
\tablecolumns{10}
\label{tab:offset-tab}
\tablehead{
\colhead{Name} & \colhead{$X_0$ East} & \colhead{$Y_0$ North} & \colhead{$R_0$} & \colhead{$R_0
/\sigma_{R_{0}}$} & \colhead{$z$} & \colhead{$D_\theta$} & \colhead{$X_0$ (proj)} & \colhead{$Y
_0$ (proj)} & \colhead{$R_0$ (proj)} \\
\colhead{} & \colhead{\arcsec} & \colhead{\arcsec} & \colhead{\arcsec} & \colhead{} & \colhead{
} & \colhead{kpc/\arcsec} & \colhead{kpc} & \colhead{kpc} & \colhead{kpc} }
 \startdata
GRB 970228
 & $-$0.033$\pm$0.034 & $-$0.424$\pm$0.034 & 0.426$\pm$0.034 &  12.59 & 0.695
 & 7.673 & $-$0.251$\pm$0.259 & $-$3.256$\pm$0.259 & 3.266$\pm$0.259 \\
 
GRB 970508
 & 0.011$\pm$0.011 & 0.001$\pm$0.012 & 0.011$\pm$0.011 & 1.003 & 0.835
 & 8.201 & 0.090$\pm$0.090 & 0.008$\pm$0.098 & 0.091$\pm$0.090 \\
 
GRB 970828
 & 0.440$\pm$0.516 & 0.177$\pm$0.447 & 0.474$\pm$0.507 & 0.936 & 0.958
 & 8.534 & 3.755$\pm$4.403 & 1.510$\pm$3.815 & 4.047$\pm$4.326 \\
 
GRB 971214
 & 0.120$\pm$0.070 & $-$0.070$\pm$0.070 & 0.139$\pm$0.070 & 1.985 & 3.418
 & 7.952 & 0.954$\pm$0.557 & $-$0.557$\pm$0.557 & 1.105$\pm$0.557 \\
 
GRB 980326
 & 0.125$\pm$0.068 & $-$0.037$\pm$0.062 & 0.130$\pm$0.068 & 1.930 & $\sim 1$
 & \ldots & \ldots & \ldots & \ldots \\
 
GRB 980329
 & $-$0.037$\pm$0.049 & $-$0.003$\pm$0.061 & 0.037$\pm$0.049 & 0.756 & $\ale 3.5$
 & \ldots & \ldots & \ldots & \ldots \\
 
GRB 980425
 & $-$10.55$\pm$0.052 & $-$6.798$\pm$0.052 &  12.55$\pm$0.052 &  241.4 & 0.008
 & 0.186 & $-$1.964$\pm$0.010 & $-$1.265$\pm$0.010 & 2.337$\pm$0.010 \\
 
GRB 980519
 & $-$0.050$\pm$0.130 & 1.100$\pm$0.100 & 1.101$\pm$0.100 &  11.00 & \ldots
 & \ldots & \ldots & \ldots & \ldots \\
 
GRB 980613
 & 0.039$\pm$0.052 & 0.080$\pm$0.080 & 0.089$\pm$0.076 & 1.174 & 1.096
 & 8.796 & 0.344$\pm$0.454 & 0.703$\pm$0.707 & 0.782$\pm$0.666 \\
 
GRB 980703$^a$
 & $-$0.054$\pm$0.055 & 0.098$\pm$0.065 & 0.112$\pm$0.063 & 1.788 & 0.966
 & 8.553 & $-$0.460$\pm$0.469 & 0.842$\pm$0.555 & 0.959$\pm$0.536 \\
 
GRB 981226
 & 0.616$\pm$0.361 & 0.426$\pm$0.246 & 0.749$\pm$0.328 & 2.282 & \ldots
 & \ldots & \ldots & \ldots & \ldots \\
 
GRB 990123
 & $-$0.192$\pm$0.003 & $-$0.641$\pm$0.003 & 0.669$\pm$0.003 &  223.0 & 1.600
 & 9.124 & $-$1.752$\pm$0.027 & $-$5.849$\pm$0.027 & 6.105$\pm$0.027 \\
 
GRB 990308
 & $-$0.328$\pm$0.357 & $-$0.989$\pm$0.357 & 1.042$\pm$0.357 & 2.919 & \ldots
 & \ldots & \ldots & \ldots & \ldots \\
 
GRB 990506
 & $-$0.246$\pm$0.432 & 0.166$\pm$0.513 & 0.297$\pm$0.459 & 0.647 & 1.310
 & 9.030 & $-$2.221$\pm$3.901 & 1.499$\pm$4.632 & 2.680$\pm$4.144 \\
 
GRB 990510
 & $-$0.064$\pm$0.009 & 0.015$\pm$0.012 & 0.066$\pm$0.009 & 7.160 & 1.619
 & 9.124 & $-$0.584$\pm$0.082 & 0.137$\pm$0.109 & 0.600$\pm$0.084 \\
 
GRB 990705
 & $-$0.865$\pm$0.046 & 0.109$\pm$0.049 & 0.872$\pm$0.046 &  18.86 & 0.840
 & 8.217 & $-$7.109$\pm$0.380 & 0.896$\pm$0.399 & 7.165$\pm$0.380 \\
 
GRB 990712
 & $-$0.035$\pm$0.080 & 0.035$\pm$0.080 & 0.049$\pm$0.080 & 0.619 & 0.434
 & 6.072 & $-$0.213$\pm$0.486 & 0.213$\pm$0.486 & 0.301$\pm$0.486 \\
 
GRB 991208
 & 0.071$\pm$0.102 & 0.183$\pm$0.096 & 0.196$\pm$0.097 & 2.016 & 0.706
 & 7.720 & 0.548$\pm$0.789 & 1.410$\pm$0.744 & 1.513$\pm$0.750 \\
 
GRB 991216
 & 0.211$\pm$0.029 & 0.290$\pm$0.034 & 0.359$\pm$0.032 &  11.08 & 1.020
 & 8.664 & 1.828$\pm$0.251 & 2.513$\pm$0.295 & 3.107$\pm$0.280 \\
 
GRB 000301C
 & $-$0.025$\pm$0.014 & $-$0.065$\pm$0.005 & 0.069$\pm$0.007 & 9.821 & 2.030
 & 9.000 & $-$0.222$\pm$0.130 & $-$0.581$\pm$0.046 & 0.622$\pm$0.063 \\
 
GRB 000418
 & $-$0.019$\pm$0.066 & 0.012$\pm$0.058 & 0.023$\pm$0.064 & 0.358 & 1.118
 & 8.829 & $-$0.170$\pm$0.584 & 0.109$\pm$0.514 & 0.202$\pm$0.564\enddata
 
\tablenotetext{a}{Using radio detections of the host and afterglow,
\citet{bkf01} find a more accurate offset of $X_0 = -0.032 \pm 0.015$
and $Y_0 = 0.025 \pm 0.015$ ($R_0 = 0.040 \pm 0.015$), consistent with
our optical results; see \S \ref{subsec:980703}.}

\tablecomments{The observed offsets ($X_0, Y_0$) and associated
Gaussian uncertainties include all statistical errors from the
astrometric mapping and OT+host centroid measurements.  The observed
offset, $R_0 = \sqrt{X_0^2 + Y_0^2}$, and $\sigma_{R_0}$ (constructed
analagously to eq.~\ref{eq:sigmar}) are given in col.~4. Note that
$R_0 - \sigma_{R_0} \le R \le R_0 - \sigma_R$ is not necessarily the
68\% percent confidence region of the true offset since the
probability distribution is not Gaussian (see eq.~\ref{eq:pr2}). The
term $R_0/\sigma_{R_0}$ in col.~5 indicates how well the offset from
the host center is determined.  In general, we consider the GRB to be
significantly displaced from the host center if $R_0/\sigma_{R_0} \age
5$. In col.~6, $z$ is the measured redshift of the host galaxy and/or
aborption redshift of the GRB complied from the literature
\citepcf{kbb+00}. In col.~7, $D_\theta$ is the conversion of
angular displacement in arcseconds to projected physical distance.  }
\end{deluxetable}
\newpage

\begin{deluxetable}{lcccccccc}
\tablecaption{Host detection probabilities and host normalized offsets\label{tab:offnorm}}
\tablecolumns{8}
\tablehead{
\colhead{Name} & \colhead{R$_{c, {\rm host}}$} & \colhead{$A_{R_c}$} & \colhead{$P_{\rm chance}$} & \colhead{$R_{\rm half}$ (obs)} & \colhead{$R_{\rm half}$ (calc)} & \colhead{$r_e$} & \colhead{$r_0$} \\
\colhead{} & \colhead{mag} & \colhead{mag} & \colhead{} & \colhead{\arcsec} & \colhead{\arcsec} & \colhead{kpc} & \colhead{} }
 \startdata

GRB 970228
 &  24.60$\pm$0.20 & 0.630 & 0.00935 &  0.345$\pm$0.030 & 0.316$\pm$0.095
 &  1.6 & 1.233$\pm$0.146 \\
 
GRB 970508
 &  24.99$\pm$0.17 & 0.130 & 0.00090 &  0.089$\pm$0.026 & 0.300$\pm$0.090
 &  0.4 & 0.124$\pm$0.129 \\
 
GRB 970828
 &  25.10$\pm$0.30 & 0.100 & 0.07037 & \ldots & 0.296$\pm$0.089
 &  1.5 & 1.603$\pm$1.780 \\
 
GRB 971214
 &  25.65$\pm$0.30 & 0.040 & 0.01119 &  0.226$\pm$0.031 & 0.273$\pm$0.082
 &  1.1 & 0.615$\pm$0.321 \\
 
GRB 980326
 &  28.70$\pm$0.30 & 0.210 & 0.01878 &  0.043$\pm$0.028 & 0.116$\pm$0.035
 &  0.2 & 3.023$\pm$2.532 \\
 
GRB 980329
 &  27.80$\pm$0.30 & 0.190 & 0.05493 &  0.245$\pm$0.033 & 0.168$\pm$0.050
 &  1.3 & 0.152$\pm$0.202 \\
 
GRB 980425
 &  14.11$\pm$0.05 & 0.170 & 0.00988 & 18.700$\pm$0.025 & \ldots
 &  2.1 & 0.671$\pm$0.003 \\
 
GRB 980519
 &  25.50$\pm$0.30 & 0.690 & 0.05213 &  0.434$\pm$0.041 & 0.279$\pm$0.084
 &  2.2 & 2.540$\pm$0.332 \\
 
GRB 980613
 &  23.58$\pm$0.10 & 0.230 & 0.00189 &  0.227$\pm$0.031 & 0.352$\pm$0.106
 &  1.2 & 0.392$\pm$0.338 \\
 
GRB 980703
 &  22.30$\pm$0.08 & 0.150 & 0.00045 &  0.169$\pm$0.026 & 0.392$\pm$0.117
 &  0.9 & 0.663$\pm$0.385 \\
 
GRB 981226
 &  24.30$\pm$0.01 & 0.060 & 0.01766 &  0.336$\pm$0.030 & 0.327$\pm$0.098
 &  1.7 & 2.227$\pm$0.996 \\
 
GRB 990123
 &  23.90$\pm$0.10 & 0.040 & 0.01418 &  0.400$\pm$0.028 & 0.341$\pm$0.102
 &  2.2 & 1.673$\pm$0.117 \\
 
GRB 990308
 &  28.00$\pm$0.50 & 0.070 & 0.31659 &  0.213$\pm$0.028 & 0.156$\pm$0.047
 &  1.1 & 4.887$\pm$1.776 \\
 
GRB 990506
 &  24.80$\pm$0.30 & 0.180 & 0.04365 &  0.090$\pm$0.027 & 0.308$\pm$0.092
 &  0.5 & 3.297$\pm$5.196 \\
 
GRB 990510
 &  27.10$\pm$0.30 & 0.530 & 0.01218 &  0.167$\pm$0.041 & 0.205$\pm$0.061
 &  0.9 & 0.393$\pm$0.111 \\
 
GRB 990705
 &  22.00$\pm$0.10 & 0.334$^{\rm a}$ & 0.01460 &  1.151$\pm$0.030 & 0.400$\pm$0.120
 &  5.7 & 0.758$\pm$0.045 \\
 
GRB 990712
 &  21.90$\pm$0.15 & 0.080 & 0.00088 &  0.282$\pm$0.026 & 0.403$\pm$0.121
 &  1.0 & 0.175$\pm$0.284 \\
 
GRB 991208
 &  24.20$\pm$0.20 & 0.040 & 0.00140 &  0.048$\pm$0.026 & 0.330$\pm$0.099
 &  0.2 & 4.083$\pm$2.994 \\
 
GRB 991216
 &  25.30$\pm$0.20 & 1.640 & 0.00860 &  0.400$\pm$0.043 & 0.288$\pm$0.086
 &  2.1 & 0.898$\pm$0.127 \\
 
GRB 000301C
 &  28.00$\pm$0.30 & 0.130 & 0.00629 &  0.066$\pm$0.028 & 0.156$\pm$0.047
 &  0.4 & 1.054$\pm$0.462 \\
 
GRB 000418
 &  23.80$\pm$0.20 & 0.080 & 0.00044 &  0.096$\pm$0.027 & 0.345$\pm$0.103
 &  0.5 & 0.239$\pm$0.670\enddata

\tablenotetext{a}{Since the GRB position pierces through the Large
Magallanic Cloud \citep{dkh+99}, we have added 0.13 mag extinction to the
Galactic extinction quoted in \citet{pia01}.  This assumes an average
extinction through the LMC of E(B-V) = 0.05 \citep{dbc+01}.}

\tablecomments{Column 2 gives the dereddened host magnitude as
referenced in \citet{pia01}, \citet{dfk+01a} and
\citet{sfc+01}. Column 3 gives the extimated extinction in the
direction of the GRB host galaxy \citet{pia01}. Column 4 gives the
estimated probability that the assigned host is a chance superposition
and not physically related to the GRB (following \S
\ref{sec:angoffs}).  The half-light radii $R_{\rm half}$ are observed
from HST imaging (col.~5) or calculated using the magnitude-radius
empirical relationship (col.~6; see text).  For HST imaging,
uncertainty is taken as the sum of the statistical error and estimated
systematics error (0\arcsec.025 which is approximately the size of one
deconvolved pixel). Otherwise, the uncertainty is taken as 30\% of the
calculated radius (col.~6). Column 7 gives the estimated host disk
scale length. The host-normalized offset $r_0 = R_0/R_{\rm half}$
given in col.~8 is derived from (if possible) the observed half-light
radius or the calculated half-light radius (otherwise).  The error on
$r_0$ is $\sigma_r$ from eq.~\ref{eq:sigmar}. Note that $r_0 -
\sigma_r \le r \le r_0 - \sigma_r$ is not necessarily the 68\% percent
confidence region of the true offset since the probability
distribution is not Gaussian.}

\end{deluxetable}

\begin{deluxetable}{lcccccccc}
\tabletypesize{\small}
\rotate
\tablewidth{9in}
\tablecaption{Comparison of observed offset distributions to various progenitor model
predictions\label{tab:nsks}} 
\tablecolumns{9}
\label{tab:nsks}
\tablehead{
\colhead{Progenitor} & \multicolumn{4}{c}{Comparison Model} & \multicolumn{3}{c}{$P_{KS}$} & \colhead{Fraction of} \\
\colhead{} & \colhead{Name} & \colhead{$r_e$} & \colhead{$v_{\rm circ}$} &
\colhead{Mass ($M_\odot$)} & \colhead{observed} & \colhead{synth} & \colhead{synth (replaced)} & \colhead{ $P_{KS} \ge 0.05$} }
\startdata

Collapsar\ldots& expon. & $R_{\rm half}/1.67$ & & & 0.454 & 0.409 & 0.401 & 0.996 \\
~~/Promptly Bursting Binary & disk  \\
NS--NS, BH--NS\ldots & a$^*$ & 1.5 kpc &  100 km s$^{-1}$ & $9.2 \times 10^9$ & 9.5 $\times 10^{-4}$  & 2.2 $\times 10^{-3}$ & 2.2 $\times 10^{-3}$  & 0.003 \\
                     & a     & 1.0 kpc & 100 km s$^{-1}$ & $9.2 \times 10^{9}$ & 6.9 $\times 10^{-3}$ & 1.7 $\times 10^{-2}$ & 1.8 $\times 10^{-2}$  & 0.139 \\
                     & b     & 1.0 kpc & 100 km s$^{-1}$ & $2.8 \times 10^{10}$ & 4.9 $\times 10^{-3}$& 2.0 $\times 10^{-2}$ &2.0 $\times 10^{-2}$  & 0.243 \\
                     & c     & 3.0 kpc & 100 km s$^{-1}$ & $2.8 \times 10^{10}$ & 3.5 $\times 10^{-5}$& 4.7 $\times 10^{-5}$ &4.7 $\times 10^{-5}$  & 0.001 \\
                     & d     & 1.0 kpc & 150 km s$^{-1}$ & $6.3 \times 10^{10}$ & 2.5 $\times 10^{-2}$& 6.0 $\times 10^{-2}$ &6.3 $\times 10^{-2}$  & 0.553 \\
                     & e     & 3.0 kpc & 150 km s$^{-1}$ & $6.3 \times 10^{10}$ & 5.7 $\times 10^{-5}$& 1.3 $\times 10^{-4}$ &1.3 $\times 10^{-4}$  & 0.001  \\
\enddata

\tablecomments{The names in column 2 correspond to the model
distribution compared against the data, with the letters corresponding
to population synthesis models from \citet{bsp99}.  For the NS--NS,
BH--NS comparisons, we believe that $a^*$ best represents
of the observed host galaxy properties (see text).  There are three
columns that give the K-S probability that the data are drawn from the
model.  The first, marked ``observed,'' is the direct comparison of
the models to the data without accounting for the uncertainties and
measurements in the data.  The second is the median K--S probability
derived from our Monte Carlo modeling (\S \ref{sec:robust}). The third
is the median K--S probability derived from our Monte Carlo modeling
but where offsets thrown out due to high $P_{\rm chance}$ are replaced
by synthetic offsets drawn from the model distribution. This pushes
$P_{\rm KS}$ to larger values, in general, but the resulting median of
the distribution is strongly affected (since $P_{\rm chance}$ is near
zero for most offsets).  The last column gives the fraction of
synthetic datasets with $P_{KS} \ge 0.05$.}

\end{deluxetable}

\end{document}